\documentclass[11pt,a4paper]{article}

\usepackage{amsmath}
\usepackage{amssymb}
\usepackage{mathrsfs}
\usepackage{amsthm}
\usepackage{graphicx}
\usepackage{color}
\usepackage{float}
\usepackage{subfigure}
\usepackage{framed}
\usepackage{ulem}
\usepackage{array}
\usepackage{makecell}
\restylefloat{table}
\usepackage[utf8]{inputenc}

\def\R{\mathbb R}

\theoremstyle{definition}
	\newtheorem{definition}{Definition}[section]
	\newtheorem{remark}{Remark}

	\newtheorem{proposition}[definition]{Proposition}
	\newtheorem{theorem}[definition]{Theorem}
\theoremstyle{remark}

	\title{General rogue wave solutions to the Sasa-Satsuma equation}
	
	\author{ Chengfa Wu$^1$,  Guangxiong Zhang$^1$,   Changyan Shi$^1$, Bao-Feng Feng$^{*2}$
\\
\\
 $^1$Institute for Advanced Study, Shenzhen University\\ Shenzhen 518060, People's Republic of China
\\
 $^2$School of Mathematical and Statistical Sciences, \\ The University of
Texas Rio Grande Valley Edinburg,
\\ Edinburg, TX 78541-2999, USA
\\
}

\date{}

\begin{document}

\maketitle

\begin{figure}[b]
\rule[-2.5truemm]{5cm}{0.1truemm}\\[2mm]
{\footnotesize  *Corresponding author. Email address: baofeng.feng@utrgv.edu.}

\end{figure}

\begin{abstract}
General rogue wave solutions to the Sasa-Satsuma equation are constructed by the Kadomtsev-Petviashvili (KP) hierarchy reduction method. These solutions are presented in three different forms. The first form is expressed in terms of recursively defined differential operators while the second form shares a similar solution structure except that the differential operators are no longer recursively defined.  Instead of using differential operators, the third form is expressed by Schur polynomials.
\end{abstract}

\section{Introduction}

The Sasa-Satsuma (SS) equation is one of the nontrivial integrable extensions of the nonlinear Schr\"odinger (NLS) equation. It was discovered by Sasa and Satsuma \cite{Sasa1991Satsuma} when they were searching integrable cases of a higher-order NLS equation proposed by Kodama and Hasegawa \cite{Kodama1985,Kodama1987Hasegawa}.
This eqeuation can be written in the form \cite{Sasa1991Satsuma}
\begin{equation} \label{Sasa1990equation4}
	\begin{aligned}
    &\mathbf{i} \frac{\partial q}{\partial T}+\frac{1}{2} \frac{\partial^{2} q}{\partial X^{2}}+|q|^{2} q+\mathbf{i} \varepsilon\left\{\frac{\partial^{3} q}{\partial X^{3}}+6|q|^{2} \frac{\partial q}{\partial X}+3 q \frac{\partial|q|^{2}}{\partial X}\right\}=0,\\
    \end{aligned}
\end{equation}
where $q$ corresponds to the complex envelope of the wave
field and the real constant $\varepsilon$  scales the integrable perturbations of the NLS equation. For the case of $\varepsilon=0$, the SS equation reduces to the NLS equation. Like the NLS equation, the SS equation contains essential terms that are commonly involved in nonlinear optics \cite{Mihalache1997TrutaCrasovan,Solli2007RopersKoonathJalali}, such as the third-order dispersion, the self-steepening and the self-frequency shift.  As a result of its integrability and physical implications, the SS equation has been studied comprehensively. Soliton solutions of the SS equation have been obtained in a number of works \cite{Mihalache1993TornerMoldoveanuPanoiuTruta,Gilson2003HietarintaNimmoOhta,Ohta2010} while the long-time asymptotic behaviour of the SS equation with decaying initial data was analyzed in \cite{Liu2018GengXue} by formulating a Riemann-Hilbert problem. Very recently, breather solutions of the SS equation were derived based on the Kadomtsev-Petviashvili (KP) hierarchy reduction method.

Rogue waves have been part of folklore for centuries in the maritime community. They refer to unusually large, unpredictable and suddenly appearing surface waves. These features indicate that they may result in tremendous danger to ships, oil platforms and coastal structures. Since the first record in 1995, rogue waves have attracted much attention from the scientific community \cite{Solli2007RopersKoonath,Dysthe2008KrogstadMuller,Chabchoub2011HoffmannAkhmediev}. 
They have been experimentally reported in diverse nonlinear systems, such as optical fibers \cite{Solli2007RopersKoonath}, Bose-Einstein condensates \cite{Bludov2009KonotopAkhmediev} and plasmas \cite{Bailung2011SharmaNakamura}. After that, there have been rapidly growing research interests in rogue waves. From the mathematical viewpoint, the first model for rogue waves is the Peregrine soliton \cite{Peregrine1983}, a particular rational solution of the NLS equation. Since the discovery of higher-order rogue waves by Akhmediev et al. \cite{akhmediev2009rogue}, there has been an explosion of research activities on the study of higher-order rogue waves. Explicit expressions of rogue wave solutions have been derived in a variety of integrable equations as well as their multi-component generalizations, such as the Davey-Stewartson equations \cite{OhtaYangDSI,OhtaYangDSII}, the Yajima-Oikawa equation \cite{chen2015rational,chen2017YORW}, a long-wave-short-wave model of Newell type \cite{CCFMYORW19}, the derivative NLS equation \cite{Yang2020ChenYang}, the three-wave equation \cite{JiankeBoIMA},  the Manakov system \cite{Baronio2014ConfortiDegasperisLombardoOnoratoWabnitz,Chen2015Mihalache} and
the coupled Hirota system \cite{Chen2013Song}. While rogue waves on the periodic background \cite{Chen2018Pelinovsky,Chen2019PelinovskyWhite,Feng2020LingTakahashi,Zhang2021Chen} have been extensively studied, rogue waves of infinite order have been revealed \cite{Bilman2020LingMiller} with the help of Riemann-Hilbert approach \cite{Yang2010}. In addition, higher-order rogue waves may exhibit universal patterns. Preliminary results on rogue wave patterns of the NLS equation were obtained in \cite{Kedziora2011AnkiewiczAkhmediev,He2013ZhangWangPorsezianFokas,Kedziora2013Ankiewicz}. Very recently, Yang and Yang \cite{Yang2021Yang-1} proved their results analytically by connecting rogue waves solutions of the NLS equation with the Yablonskii-Vorob'ev polynomial hierarchy. Moreover, they \cite{Yang2021Yang-2} showed that similar rogue wave patterns also appear in many other integrable equations.

Rogue wave solutions of the NLS equation have been investigated comprehensively and numerous satisfactory results have been achieved. Therefore, it is natural to extend these results to the SS equation that serves as an integrable extension of the NLS equation. Fundamental rogue wave solutions of the SS equation were first derived by Bandelow and Akhmediev \cite{Bandelow2012Akhmediev-1,Bandelow2012Akhmediev} by taking the long wave limit of breather solutions. With the help of Darboux dressing method, Chen \cite{Chen2013} revealed a distinctive characteristic of the fundamental rogue wave of the SS equation, which is the so-called twisted-rogue wave pair. Explicit second order rogue waves were obtained by Mu et al. \cite{MU2020QinRogerNail} and their structure is much more complicated than the case of the NLS equation. Higher-order rogue waves were considered in  \cite{Ling2016,Mu2016Qin}. Nevertheless, explicit forms of higher-order rogue waves of the SS equation are still lack of exploring.

The main objective of this paper is to construct general rogue wave solutions to the Sasa-Satsuma equation \cite{Ohta2010,Wu2022WeiShiFeng}
\begin{equation}\label{SS equation}
u_{t}=u_{x x x}-6 c|u|^{2} u_{x}-3 c u\left(|u|^{2}\right)_{x},
\end{equation}
where $c$ is a real constant. This form is more convenient for our consideration and for the case of $c=-1$, it can be obtained from \eqref{Sasa1990equation4} via Gauge, Galilean and scale transformations. We will adopt the Kadomtsev-Petviashvili (KP) hierarchy reduction method. This is a very powerful method to construct explicit solutions of integrable equations and has found extensive applications in the derivation of rogue wave or soliton solutions of various integrable equations \cite{Ohta2012Yang,OhtaYangDSI,OhtaYangDSII,chen2017YORW,Yang2020ChenYang,JiankeBoIMA}. However, several difficulties are encountered when applying this method to derive rogue wave solutions to the Sasa-Satsuma equation due to its complexity.
 Unlike the NLS equation which belongs to the AKP hierarchy, the SS equation belongs to the CKP hierarchy which is a symmetry reduction and the sub-hierarchy of the AKP hierarchy \cite{JM83}. This forces us to start with a kernel of $2 \times 2$ matrix which is one of the novelties of the present paper. On the other hand, the bilinear equations of the Sasa-Satsuma equation correspond to eleven bilinear equations in the KP hierarchy. Therefore, as explained in subsequent sections, the reduction procedure and the rogue wave solutions are much more complicated.

 The structure of this paper is as follows. In Section \ref{General rogue wave solutions to the Sasa-Satsuma equation}, we present the general rogue wave solutions to the Sasa-Satsuma equation. These solutions are given in three different forms. The first two forms are expressed in terms of differential operators. The difference between them is that the differential operators are recursively defined for one form whereas they are no longer recursively defined for the other form.
  Instead of using differential operators, the third form is represented by Schur polynomials. Then the solution dynamics are briefly examined in Section \ref{Dynamics of rogue wave solutions} and the proofs of our main results are provided in Sections \ref{Proof of Theorem-recursive relation}-\ref{Proof of Theorem-Schur polynomials}. Finally, we conclude this paper in Section \ref{Conclusions}.

\section{General rogue wave solutions to the Sasa-Satsuma equation}\label{General rogue wave solutions to the Sasa-Satsuma equation}
In this section, we present the general rogue wave solutions to the SSE \eqref{SS equation}. These solutions are presented in three different forms.
The first two forms are expressed in terms of differential operators while the third form is expressed by Schur polynomials.
The difference between the first two forms is that the differential operators are recursively defined in one of them whereas the other one does not involve recursive relations among the  differential operators.

\begin{theorem}\label{thm-recursive relation}
Let $a= i \kappa$, where $\kappa$ is real,  $ F(p)$ be a function defined by
\begin{equation} \label{definition of F}
  F(p) =  \dfrac{1}{p-a} + \dfrac{1}{p+a} - \dfrac{p}{c}
\end{equation}
and $\xi$ be a root of
 \begin{equation} \label{parapeter value p}
  F'(p) = - \dfrac{1}{(p-a)^2} - \dfrac{1}{(p+a)^2} - \dfrac{1}{c} = 0.
 \end{equation}
Then the Sasa-Satsuma  equation \eqref{SS equation}
admits the general rogue wave solutions 
\begin{equation}\label{rogue wave solution-recursive relation}
 u=\frac{g}{f} e^{\mathrm{i}\left(\kappa(x-6 c t)-\kappa^{3} t\right)}
\end{equation}
where 
$$f(x,t)=\tau_{0} (x-6 c t,t), \quad g(x,t)=\tau_{1} (x-6 c t,t)$$ and $\tau_{k} \, (k=0,1)$ is defined as
\begin{eqnarray*}
\tau_{k} =
 \det_{1 \leq i, j \leq N} \left(M_{ij}^k\right)   
\end{eqnarray*}
where 
\begin{eqnarray*}
M_{ij}^k &=&\left. \left(
\begin{array}{cccc}
   A_{2i-1}^{(N-i)}(p_1) B_{2j-1}^{(N-j)}(q_1) m_{11}^{k} &  A_{2i-1}^{(N-i)}(p_1) B_{2j-1}^{(N-j)}(q_2) m_{12}^{k}   \\
 A_{2i-1}^{(N-i)}(p_2) B_{2j-1}^{(N-j)}(q_1) m_{21}^{k} &  A_{2i-1}^{(N-i)}(p_2) B_{2j-1}^{(N-j)}(q_2) m_{22}^{k}   \\
\end{array}
\right)\right|_{\substack{p_1= q_1=\xi, \\ p_2=q_2= \xi^*}}
\end{eqnarray*}
with $(\alpha,\beta =1,2,\mu,\nu=0,1,\cdots, N-1)$
\begin{eqnarray*}
m_{\alpha \beta}^{k}&=&\frac{1}{p_{\alpha}+q_{\beta}}\left(-\dfrac{p_{\alpha}-a}{q_{\beta}+a}\right)^{k} e^{\xi_{\alpha}+\eta_{\beta}}  \\
\xi_{\alpha}&=&p_{\alpha} x+p_{\alpha}^{3} t+\xi_{\alpha 0}, \quad \eta_{\beta} =  q_{\beta} x+q_{\beta}^{3} t+\eta_{\beta 0}
\\
A_{2i-1}^{(\mu)}(p) &=& \sum_{n=0}^{2i-1} \dfrac{a_{2i-1-n}^{(\mu)}(p)}{n!}  (p\partial_p)^{n}, \quad B_{2j-1}^{(\nu)}(q) = \sum_{n=0}^{2j-1} \dfrac{b_{2j-1-n}^{(\nu)}(q)}{n!}  (q\partial_q)^{n}
\\
a_k^{(\mu+1)}(p) &=& \sum_{n=0}^k \dfrac{(p\partial_p)^{n+2}F(p)}{(n+2)!}  a_{k-n}^{(\mu)}(p), \quad
   b_k^{(\nu+1)}(q) = \sum_{n=0}^k \dfrac{ (q\partial_q)^{n+2}F(q)}{(n+2)!}  b_{k-n}^{(\nu)}(q),
  \end{eqnarray*}
Here, the parameters $  p_{j}, q_{j}, \xi_{j0}, \eta_{j0}, a_n^{(0)}, b_n^{(0)} \, (j=1,2, n=0,1,\dots,2N-1)$  satisfy the constraints
 \begin{eqnarray} \label{parameter relation-recursive relation}
 p_j = q_j, \quad  p_{j} = p_{3-j}^{*}, \quad    \xi_{j0} = \eta_{j0},    \quad     \xi_{j}  =   \xi_{3-j, 0}^*, \quad
 a_n^{(0)} (p_j) = b_n^{(0)} (q_j), \quad a_n^{(0)} (p_j) = [a_n^{(0)} (p_{3-j})]^*.
 \end{eqnarray}

\end{theorem}

\begin{remark} \label{root sturcture}

We note that the algebraic equation \eqref{parapeter value p} can be rewritten as
\begin{equation} \label{parameter p}
p^4 + 2 \left( c+ \kappa ^2\right)p^2 + \kappa ^2(\kappa^2-2 c) = 0.
\end{equation}
It is also easy to see that the root $\xi$ of the above equation can be neither real nor purely imaginary otherwise non-trivial rogue wave solutions \eqref{rogue wave solution-recursive relation} do not exist.

 Further, the equation \eqref{parameter p} has at least one pair of complex conjugate roots $(\xi, \xi^*)$ with $\xi^2 \not \in \R$ only when  $c \left(c+4 \kappa ^2\right) <0$. Hence,  the rogue wave solution \eqref{rogue wave solution-recursive relation} exists if and only if $c$ and $\kappa$ satisfy the conditions
 \begin{equation}\label{paramenter constraint for the existence of RW solutions}
   c < 0, \quad c + 4 \kappa ^2>0.
 \end{equation}
On the other hand, as \eqref{parameter p} is a quartic equation in $\xi$, it has four roots (counting multiplicity). Due to the fact that all coefficients of \eqref{parameter p} are real, these roots demonstrate a symmetric structure when $c$ and $\kappa$ satisfy \eqref{paramenter constraint for the existence of RW solutions}, that is, if $\xi$ is a root of \eqref{parameter p}, then the other three roots are $-\xi$ and $\pm\xi^*$. As a result,  they can be explicitly expressed as
\begin{eqnarray*}
 \pm \dfrac{1}{\sqrt{2}}   \left[ \left(|\kappa |  (\kappa ^2-2c)^{1/2}- c-\kappa ^2\right)^{1/2}
  \pm i \left(|\kappa |  (\kappa ^2-2c)^{1/2}+ c+\kappa ^2\right)^{1/2}\right].
\end{eqnarray*}
Apart from this, the symmetric structure indicates that all the roots of $F'(p) = 0$ are simple.
\end{remark}
In Theorem \ref{thm-recursive relation}, we use recursive relations to define the differential operators that appear in the matrix elements of the $\tau$ function. To avoid these recursive relations,  Yang and Yang have proposed the so-called `$\mathcal{W}$-$p$ treatment' method which has been applied to derive rogue wave solutions of the Boussinesq equation \cite{Yang2020Yang} and the three-wave equations \cite{Yang2021Yang-1}. The key point of this method is to introduce more general differential operators in the $\tau$ function. It turns out that this method applies to the Sasa-Satsuma equation \eqref{SS equation} as well and the result is give as follows.

\begin{theorem} \label{thm-no recursive relation}
The Sasa-Satsuma  equation \eqref{SS equation}
admits the general rogue wave solutions 
\begin{equation}\label{rogue wave solution-no recursive relation}
 u=\frac{g}{f} e^{\mathrm{i}\left(\kappa(x-6 c t)-\kappa^{3} t\right)}
\end{equation}
where   $\kappa$ is real,
$$f(x,t)=\tau_{0} (x-6 c t,t), \quad g(x,t)=\tau_{1} (x-6 c t,t)$$ and $\tau_{k} \, (k=0,1)$ is defined as
\begin{eqnarray*}
\tau_{k} =
\det_{1 \leq i,j \leq N}\left(\mathcal{M}_{2i-1,2j-1}^k\right)
\end{eqnarray*}
where 
\begin{eqnarray*}
\mathcal{M}_{ij}^k = \left.  \left(
\begin{array}{cccc}
   \mathcal{A}_{i1} \mathcal{B}_{j1} m_{11}^{k} &  \mathcal{A}_{i1}  \mathcal{B}_{j2}  m_{12}^{k}   \\
 \mathcal{A}_{i2} \mathcal{B}_{j1}  m_{21}^{k} &  \mathcal{A}_{i2} \mathcal{B}_{j2}  m_{22}^{k}   \\
\end{array}
\right)
\right|_{p_1=q_1=\xi, p_2=q_2= \xi^*}
\end{eqnarray*}
with $(\alpha, \beta =1,2)$
\begin{equation}\label{differential operator-without recursive relation}
   \mathcal{A}_{i\alpha}=\frac{1}{i !}\left[f_{1\alpha}(p_{\alpha}) \partial_{p_{\alpha}}\right]^{i}, \quad \mathcal{B}_{j\beta}=\frac{1}{j !}\left[f_{2\beta}(q_{\beta}) \partial_{q_{\beta}}\right]^{j}.
\end{equation}
Here
\begin{equation} 
  f_{\mathfrak{i}1}(z)=\pm \frac{\sqrt{F^{2}(z)-F^{2}\left(\xi\right)}}{F^{\prime}(z)}, \quad  f_{\mathfrak{i}2}(z)=\pm \frac{\sqrt{F^{2}(z)-F^{2}\left(\xi^*\right)}}{F^{\prime}(z)}, \quad \mathfrak{i} =1,2.
\end{equation}
All other functions and parameters are the same as in Theorem \ref{thm-recursive relation} except that
\begin{eqnarray*}
  \xi_{10} = \sum_{n=1}^{\infty}  a_{n} \ln ^{n} \mathcal{U}_{1}(p_1), \quad  \eta_{10}= \sum_{n=1}^{\infty}  a_{n}  \ln ^{n} \mathcal{V}_{1}(q_1),
\end{eqnarray*}
where
\begin{equation}
 \mathcal{U}_{1}(p_1)=\frac{F(p_1) \pm \sqrt{F^{2}(p_1)-F^{2}\left(\xi\right)}}{F\left(\xi\right)}, \quad \mathcal{V}_{1}(q_1)=\frac{F(q_1) \pm \sqrt{F^{2}(q_1)-F^{2}\left(\xi\right)}}{F\left(\xi\right)}
\end{equation}
and $a_{r}$ are free complex constants.

\end{theorem}

Finally, we show that the solutions obtained in Theorem \ref{thm-no recursive relation} can be converted into a more explicit manner by means of Schur polynomials.
These Schur polynomials are defined via the generating function
	$$
	\sum_{n=0}^{\infty}S_n(\textbf{x})\lambda^n=\exp\left(\sum_{k=1}^{\infty}x_k\lambda^k\right),
	$$
	where $\textbf{x}=(x_1,x_2,\cdots)$. To be more explicit, we have
\begin{equation}\label{Schur polynomials}
  S_{0}(\mathbf{x})=1, \quad S_{1}(\mathbf{x})=x_{1}, \quad S_{2}(\mathbf{x})=\frac{1}{2} x_{1}^{2}+x_{2},  \ldots, \quad S_{j}(\mathbf{x})=\sum_{l_{1}+2 l_{2}+\cdots+m l_{m}=j}\left(\prod_{i=1}^{m} \frac{x_{i}^{l_{i}}}{l_{i} !}\right).
\end{equation}

\begin{theorem} \label{thm-Schur polynomials}
Let $a= i \kappa$, where $\kappa$ is real,  $\mathfrak{p}(\kappa)$ be the function defined by
\begin{equation}
  F[\mathfrak{p}(\kappa)] = F(\xi) \cosh(\kappa),
\end{equation}
and $\hat{p}_1 = \mathfrak{p}'(0)$, where $F(p)$ and $\xi$ are the same as in Theorem \ref{thm-recursive relation} and $\mathfrak{p}(0) = \xi$. Further, let  $\mathbf{x} (k) = (x_1,x_2,\cdots), \mathbf{h}^{\pm} = (h^{\pm}_1,h^{\pm}_2,\cdots), \mathbf{s}  = (s_1,s_2,\cdots), \mathbf{\hat{s}}  = (\hat{s}_1,\hat{s}_2,\cdots),  \mathbf{b}  = (b_1,b_2,\cdots)$ and $\mathbf{\hat{b}}  = (\hat{b}_1,\hat{b}_2,\cdots)$ be the vectors defined by
\begin{eqnarray}
&& x_{n} (k) = c_n x + d_n t  + a_n,
\\
&&\sum_{n=1}^{\infty} b_n \kappa^n=
   \ln \left[ \dfrac{\mathfrak{p}(\kappa)+\xi}{2 \xi}\right],  \quad
  \sum_{n=1}^{\infty} \hat{b}_n \kappa^n=
   \ln \left[ \dfrac{\mathfrak{p}(\kappa)+\xi^*}{\xi+\xi^*}\right]
   \\
&&\sum_{n=1}^{\infty} s_n \kappa^n = \ln \left[ \frac{\mathfrak{p}(\kappa)-\xi }{\hat{p}_1 \kappa} \dfrac{2\xi}{\mathfrak{p}(\kappa)+\xi}\right], \quad
  \sum_{n=1}^{\infty} \hat{s}_n \kappa^n=
   \ln \left[\frac{ \mathfrak{p}(\kappa)-\xi }{\hat{p}_{1} \kappa} \frac{\xi+\xi^*}{\mathfrak{p}(\kappa)+\xi^*}\right]
   \\
&& \sum_{n=1}^{\infty} h^+_n \kappa^n = \ln    \left[\dfrac{\mathfrak{p} (\kappa)-a}{\xi-a}\right], \quad \sum_{n=1}^{\infty} h^-_n \kappa^n = \ln    \left[\dfrac{\mathfrak{p} (\kappa)+a}{\xi+a}\right].
\end{eqnarray}
where $a_n$ is arbitrary, and  $c_n, d_n$ are defined by the expansions
\begin{eqnarray}
\sum_{n=1}^{\infty} c_n \kappa^n =  \mathfrak{p} (\kappa)- \xi   , \quad  \sum_{n=1}^{\infty} d_n \kappa^n =  \mathfrak{p^3} (\kappa)- \xi^3.
\end{eqnarray}
Then the Sasa-Satsuma  equation \eqref{SS equation}
admits the general rogue wave solutions 
\begin{equation}\label{rogue wave solution-Schur polynomial}
 u=\frac{g}{f} e^{\mathrm{i}\left(\kappa(x-6 c t)-\kappa^{3} t\right)}
\end{equation}
where  
$$f(x,t)=\sigma_{0} (x-6 c t,t), \quad g(x,t)=\sigma_{1} (x-6 c t,t)$$ and $\sigma_{k} \, (k=0,1)$ is defined as
\begin{eqnarray*}
\sigma_{k}  =
\det_{1 \leq i,j \leq N}\left(\mathcal{M}_{2i-1,2j-1}^k\right)
\end{eqnarray*}
with 
\begin{eqnarray*}
\mathcal{M}_{ij}^k =  \left( \begin{array}{ll}
  \mathfrak{m}_{2i-1,2j-1}^{k} & \mathfrak{m}_{2i-1,2j}^{k} \\
    \mathfrak{m}_{2i,2j-1}^{k} & \mathfrak{m}_{2i,2j}^{k} \\
  \end{array}
  \right)
\end{eqnarray*}
and 
 \begin{eqnarray*}
   \mathfrak{m}_{2i-1,2j-1}^{k} &=&\sum_{\nu=0}^{\min (i, j)} \frac{1}{2\xi} \left(\frac{\hat{p}_{1}^2}{4\xi^{2}}\right)^{\nu} S_{i-\nu}\left(\mathbf{x}+\nu \mathbf{s} - \mathbf{b} + k\mathbf{h}^+\right) S_{j-\nu}\left(\mathbf{x}+\nu \mathbf{s} - \mathbf{b} - k\mathbf{h}^-\right)
    \\
      \mathfrak{m}_{2i-1,2j}^{k}  &=& \sum_{\nu=0}^{\min (i, j)} \frac{1}{\xi+\xi^*} \left(\frac{|\hat{p}_{1}|^2}{(\xi+\xi^*)^{2}}\right)^{\nu}
 S_{i-\nu}\left(\mathbf{x}+\nu \mathbf{\hat{s}}- \mathbf{\hat{b}}+ k \mathbf{h}^{+}\right) S_{j-\nu}\left(\mathbf{x}^{*}+\nu \mathbf{\hat{s}}^*- \mathbf{\hat{b}}^*-k(\mathbf{h}^+)^* \right) \nonumber
       \\
      \mathfrak{m}_{2i,2j-1}^{k} &=& \sum_{\nu=0}^{\min (i, j)} \frac{1}{\xi^*+\xi} \left(\frac{|\hat{p}_{1}|^2}{(\xi^*+\xi)^{2}}\right)^{\nu}
S_{i-\nu}\left(\mathbf{x}^*+\nu \mathbf{\hat{s}}^*- \mathbf{\hat{b}^*}+k (\mathbf{h}^{-})^*\right) S_{j-\nu}\left(\mathbf{x}+\nu \mathbf{\hat{s}}- \mathbf{\hat{b}}-k\mathbf{h}^-\right) \nonumber
       \\
      \mathfrak{m}_{2i,2j}^{k} &=& \sum_{\nu=0}^{\min (i, j)} \frac{1}{2 \xi^*} \left(\frac{\left(\hat{p}_{1}^*\right)^2}{4 (\xi^*)^{2}}\right)^{\nu}
 S_{i-\nu}\left(\mathbf{x}^*+\nu \mathbf{s}^*- \mathbf{b}^*+k (\mathbf{h^-})^*\right) S_{j-\nu}\left(\mathbf{x}^*+\nu \mathbf{s}^* - \mathbf{b}^*-k(\mathbf{h^+})^* \right). \nonumber
 \end{eqnarray*}

\end{theorem}

\begin{remark}
The function $\sigma_k \, (k=0,1)$ in Theorem \ref{thm-Schur polynomials} is represented by a $N\times N$ block matrix where each block is a $2\times 2$ matrix. In fact, by applying row and column operations, we are able to rewrite $\sigma_k$ as a $2\times 2$ block matrix where each block is a $N\times N$  matrix, that is
\begin{equation*}
\sigma_{k}=\operatorname{det}\left(\begin{array}{cc}
\sigma_{k}^{[1,1]} & \sigma_{k}^{[1,2]} \\
\sigma_{k}^{[2,1]} & \sigma_{k}^{[2,2]}
\end{array}\right),
\end{equation*}
where
\begin{equation*}
\sigma_{k}^{[1,1]}=\left(\mathfrak{m}_{2i-1,2j-1}^{k}\right), \quad \sigma_{k}^{[1,2]}=\left(\mathfrak{m}_{2i-1,2j}^{k}\right), \quad \sigma_{k}^{[2,1]}=\left(\mathfrak{m}_{2i,2j-1}^{k}\right), \quad \sigma_{k}^{[2,2]}=\left(\mathfrak{m}_{2i,2j}^{k}\right),
\end{equation*}
and $\mathfrak{m}_{I,J} \, (1 \leq I,J \leq 2N)$ are defined in Theorem \ref{thm-Schur polynomials}. In a similar way, we may express the $\tau$-function in Theorems 
\ref{thm-recursive relation} and \ref{thm-no recursive relation} as a $2\times 2$ block matrix with each block being a $N\times N$  matrix.

\end{remark}

\begin{remark}
The function $\sigma_k \, (k=0,1)$ in Theorem \ref{thm-Schur polynomials} is a polynomial in $x$ and $t$. It can be computed that for the $N$-th order rogue wave, the degree of $\sigma_k (k=0,1)$ is $2N(N+1)$ for both variables $x$ and $t$. Since the computations are very similar to those developed by Yang and Yang (see Appendix A in \cite{Yang2021Yang-3}), we omit the details.

\end{remark}

%
%

\begin{remark}
The rogue wave solutions of order $N$ presented in Theorem \ref{thm-Schur polynomials} contain $2N-1$ free parameters $a_{1}, a_{2}, \ldots, a_{2 N-1}$. By a shift of $x$ and $t$, we may normalize $a_1=0$. Using similar arguments as in \cite{Ohta2012Yang,Yang2020ChenYang}, it can be shown that these rogue wave solutions are independent of the parameters $a_{2n}$, where $n$ is a positive integer. As a consequence, we will set $a_{2},a_{4},\cdots,a_{2n},\cdots$ to be $0$ in the subsequent discussions.
\end{remark}

\section{Dynamics of rogue wave solutions}\label{Dynamics of rogue wave solutions}
In this section, we analyze the dynamics of rogue wave solutions of the Sasa-Satsuma equation derived in
Theorem \ref{thm-Schur polynomials}.

\begin{figure}[h]
    \centering
\subfigure[]{%
        \includegraphics[width=150mm]{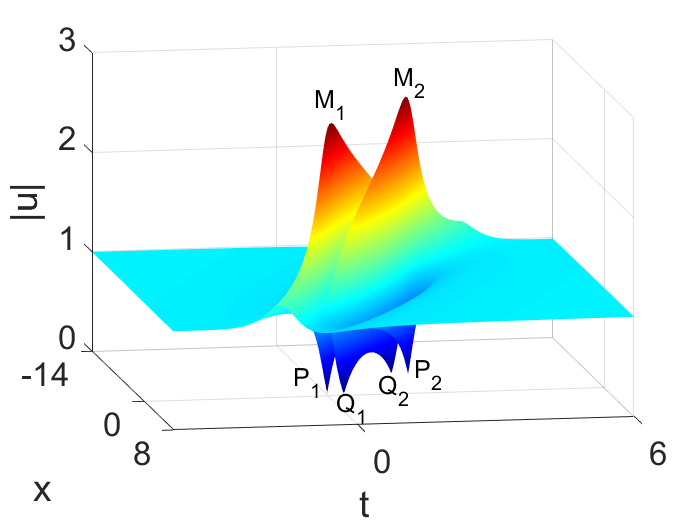}}
    \caption{(Color online) The twisted rouge wave pair with parameter values \(a=0.45\mathrm{i},c=-0.5\). }
    \label{fig:twisted RW}
\end{figure}

\begin{figure}[h]
    \centering
\subfigure[]{%
        \includegraphics[width=50mm]{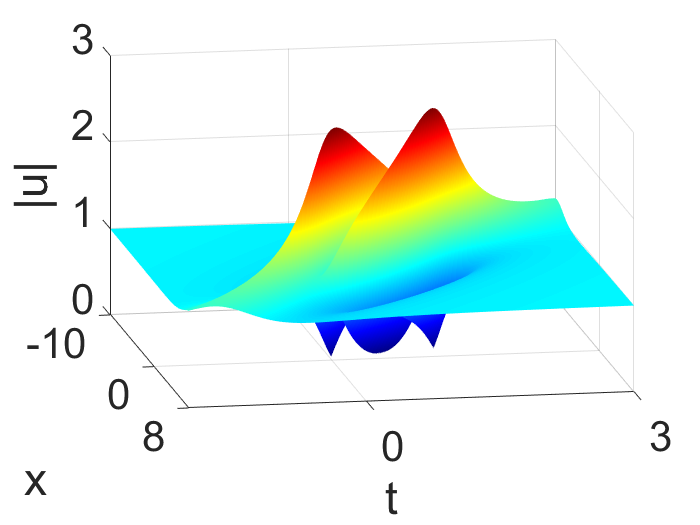}
        \label{fig:figure3(a)}}
\subfigure[]{%
        \includegraphics[width=50mm]{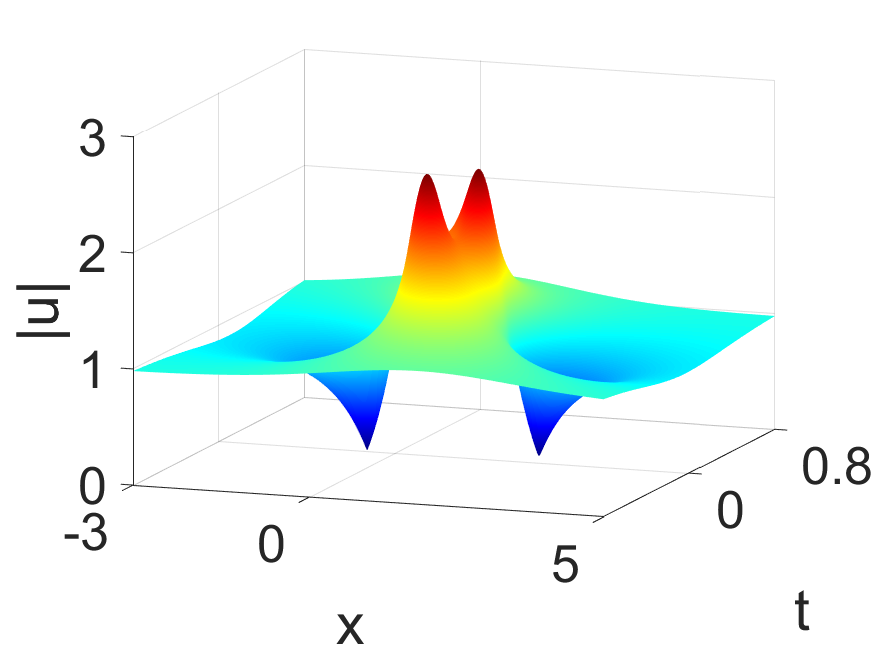}
        \label{fig:figure3(b)}}
\subfigure[]{%
        \includegraphics[width=50mm]{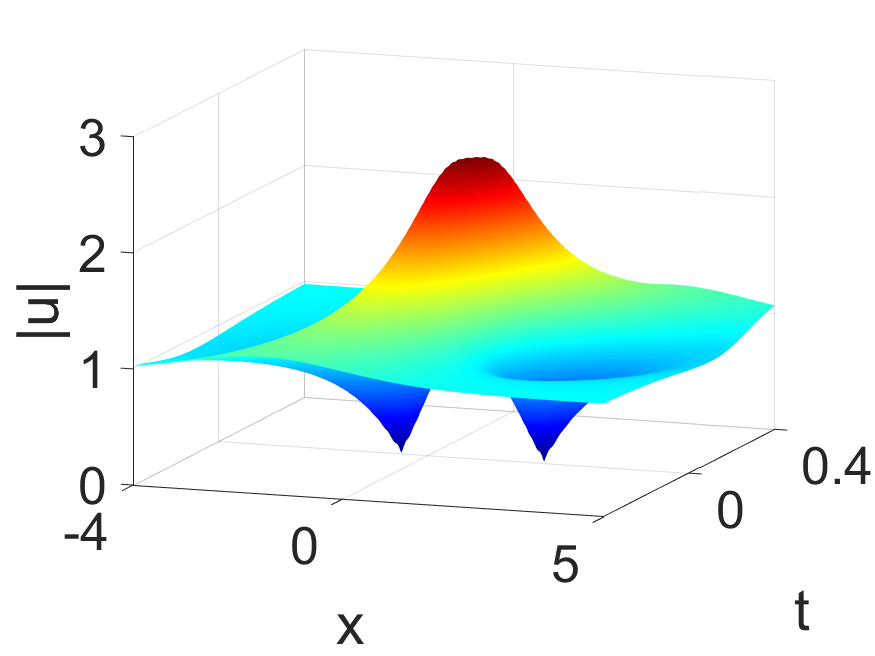}
        \label{fig:figure3(c)}}
\subfigure[]{%
        \includegraphics[width=50mm]{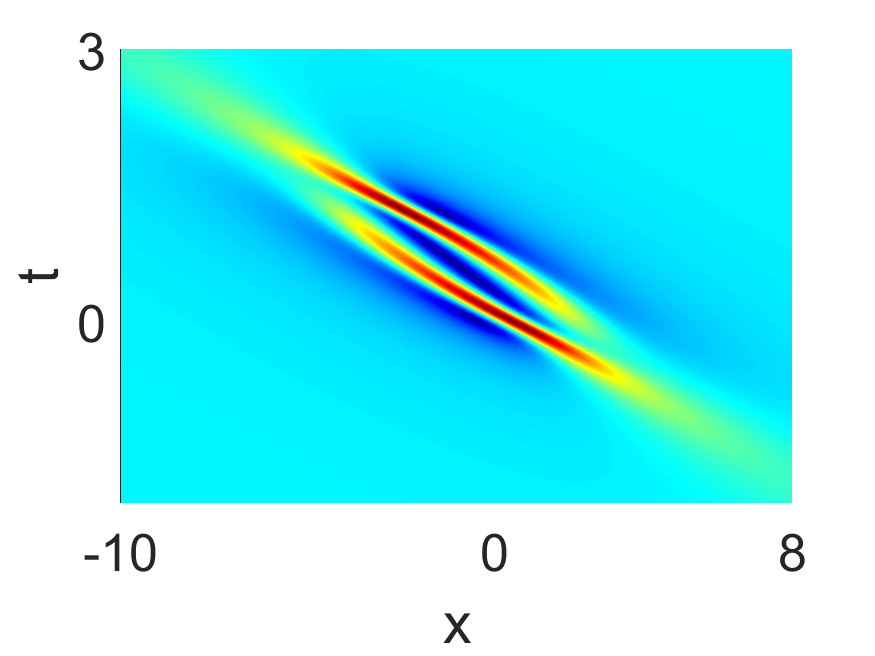}
        \label{fig:figure3(d)}}
\subfigure[]{%
        \includegraphics[width=50mm]{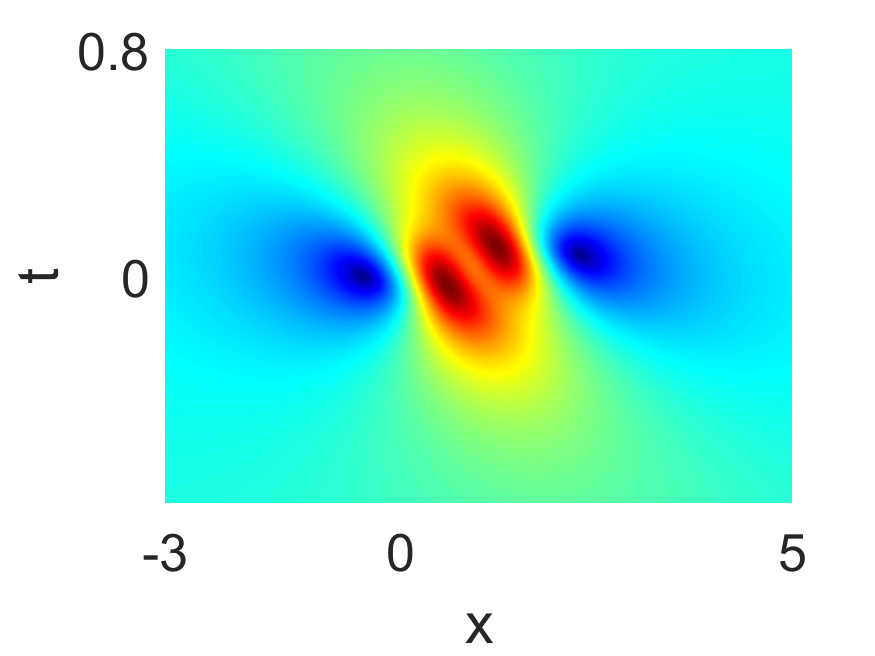}
        \label{fig:figure3(e)}}
\subfigure[]{%
        \includegraphics[width=50mm]{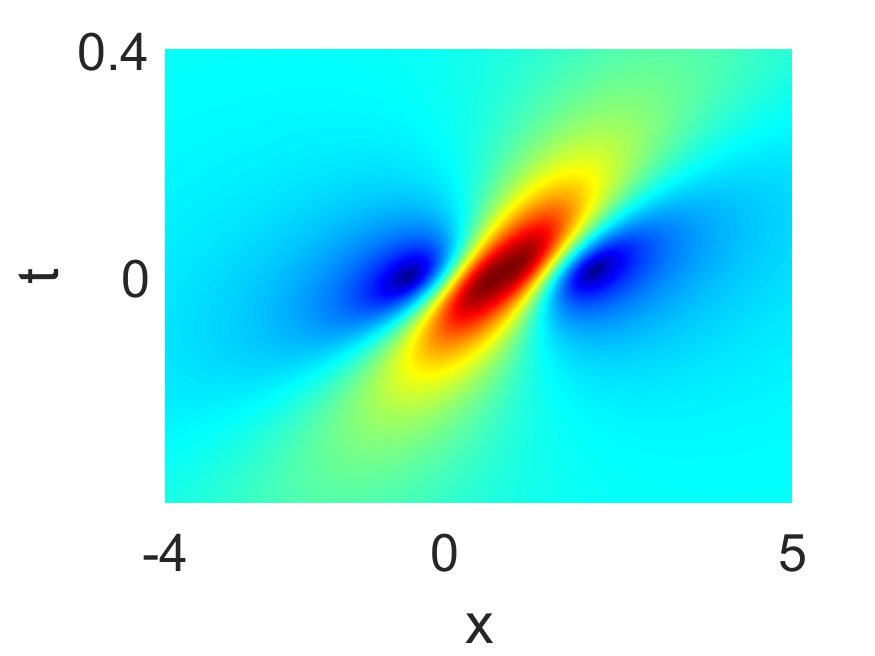}
        \label{fig:figure3(f)}}
    \caption{(Color online) First-order rouge waves under parameter values \(c=-0.5,a_1=0\) (a) \(a=0.5\mathrm{i}\), (c) \(a=1.04\mathrm{i}\) and (e) \(a=1.84\mathrm{i}\). (d), (e) and (f) are the corresponding density plots of (a), (b) and (c) respectively.}
    \label{fig:first-order RW}
\end{figure}

To obtain the fundamental rogue wave solutions, we set $N = 1$ in Theorem \ref{thm-Schur polynomials}.
In this case, we have
\begin{eqnarray*}
\sigma_0 = \left|
\begin{array}{cc}
   \mathfrak{m}_{11}^{(0)} &\mathfrak{m}_{12}^{(0)}   \\
 \mathfrak{m}_{21}^{(0)} &\mathfrak{m}_{22}^{(0)}
\end{array}
\right|
,
\quad
\sigma_1 = \left|
\begin{array}{cc}
   \mathfrak{m}_{11}^{(1)} &\mathfrak{m}_{12}^{(1)}   \\
 \mathfrak{m}_{21}^{(1)} &\mathfrak{m}_{22}^{(1)}
\end{array}
\right|,
\end{eqnarray*}
where ($k=0,1$)
\begin{eqnarray*}
  \mathfrak{m}_{11}^{(k)} &=& \frac{1}{2 \xi} \left[ (c_1 x + d_1 t + a_1 - b_1 + k h_1^+) (c_1 x + d_1 t + a_1 - b_1 - k h_1^-) + \left(\frac{\hat{p}_1}{2\xi}\right)^2 \right]  \\
  \mathfrak{m}_{12}^{(k)} &=&  \frac{1}{\xi + \xi^*} \left[ (c_1 x + d_1 t + a_1 - \hat{b}_1 + k h_1^+) (c_1^* x + d_1^* t + a_1^* - \hat{b}_1^* - k (h_1^+)^*) + \frac{|\hat{p}_1|^2}{(\xi+\xi^*)^2} \right]  \\
  \mathfrak{m}_{21}^{(k)} &=&  \frac{1}{\xi + \xi^*} \left[ (c_1^* x + d_1^* t + a_1^* - \hat{b}_1^* + k (h_1^-)^*) (c_1 x + d_1 t + a_1 - \hat{b}_1 - k h_1^-) + \frac{|\hat{p}_1|^2}{(\xi+\xi^*)^2} \right]  \\
  \mathfrak{m}_{22}^{(k)} &=&  \frac{1}{2 \xi^*} \left[  (c_1^* x + d_1^* t + a_1^* - b_1^* + k (h_1^-)^*) (c_1^* x + d_1^* t + a_1^* - b_1^* - k (h_1^+)^*) + \left(\frac{\hat{p}_1^*}{2\xi^*}\right)^2  \right]
\end{eqnarray*}
and
\begin{eqnarray*}
\hat{p}_1= \pm  \sqrt{\frac{F(\xi) \left(\xi^2-a^2\right)^3}{4 \left(3 a^2 \xi+\xi^3\right)}}, \quad b_1= \frac{\hat{p}_1}{2 \xi },   \quad \hat{b}_1 = \frac{\hat{p}_1}{\xi +\xi^*},   \quad c_1 =\hat{p}_1,   \quad d_1 = 3 \xi ^2 \hat{p}_1,   \quad h_1^+ = \frac{\hat{p}_1}{\xi -a },   \quad h_1^-= \frac{\hat{p}_1}{\xi +a }.
\end{eqnarray*}
Without loss of generality, we may set $a_1 = 0$ by a shift of $x$ and $t$. Then the fundamental rogue wave solutions of the Sasa-Satsuma equation \eqref{SS equation} are given by
\begin{equation}\label{fundamental rogue wave solution-Schur polynomial}
 u=\frac{g}{f} e^{\mathrm{i}\left(\kappa(x-6 c t)-\kappa^{3} t\right)},
\end{equation}
where  
\begin{eqnarray*}
  f(x,t) &=&  \left| \frac{1}{2\xi} \left[ \left(x + 3 (\xi^2-2c) t -\frac{1}{2 \xi}\right)^2 + \frac{1}{4 \xi^2} \right] \right|^2 -  \frac{1}{(\xi + \xi^*)^2} \left[ \left|x+ 3(\xi^2-2c) t -\frac{1}{ \xi + \xi^*}\right|^2 + \frac{1}{(\xi + \xi^*)^2} \right]^2\\
  g(x,t) &=& \frac{1}{4|\xi|^2}    \left[ \left(x + 3 (\xi^2-2c) t -\frac{1}{2 \xi} +\frac{1}{ \xi- i \kappa} \right) \left(x + 3 (\xi^2-2c) t -\frac{1}{2 \xi} -\frac{1}{ \xi+ i \kappa} \right) + \frac{1}{4\xi^2} \right]
  \\
   &&\left[ \left(x + 3 ((\xi^*)^2-2c) t -\frac{1}{2 \xi^*} +\frac{1}{ \xi^*- i \kappa} \right) \left(x + 3 ((\xi^*)^2-2c) t -\frac{1}{2 \xi^*} -\frac{1}{ \xi^*+ i \kappa} \right) + \frac{1}{4 (\xi^*)^2} \right]
   \\
   &-&\! \frac{1}{(\xi + \xi^*)^2}\! \left[\!  \left(\!x + 3 (\xi^2-2c) t -\frac{1}{\xi + \xi^*} +\frac{1}{ \xi - i \kappa} \!\right)\! \left(\!x + 3 ((\xi^*)^2-2c) t -\frac{1}{\xi + \xi^*}\! -\!\frac{1}{ \xi^* + i \kappa} \!\right)\! +\! \frac{1}{(\xi+\xi^*)^2}\! \right]
   \\
   && \left[  \left(x + 3 ((\xi^*)^2-2c) t -\frac{1}{\xi + \xi^*} +\frac{1}{ \xi^* - i \kappa} \right) \left(x + 3 (\xi^2-2c) t -\frac{1}{\xi + \xi^*} -\frac{1}{ \xi + i \kappa} \right) + \frac{1}{(\xi+\xi^*)^2} \right]
\end{eqnarray*}
We note that the above solutions are the same as the fundamental rogue wave solutions derived in Theorem \ref{thm-recursive relation} up to multiplication of a constant with unit modulus. It is clear that the solution set of \eqref{SS equation} is invariant under multiplication of $e^{i\theta}$, where $\theta$ is real, so this is reasonable. Since the dynamics of the fundamental rogue wave solutions corresponding to Theorem \ref{thm-recursive relation} have been thoroughly examined in \cite{Feng2022ShiZhangWu}, we will omit the details. Instead, we only recall some distinctive features. On the one hand, the $\tau$-function of the fundamental rogue wave solutions \eqref{fundamental rogue wave solution-Schur polynomial} is a polynomial of degree $4$ in $x$ and $t$ while it is degree $2$ for other integrable systems. Therefore, the patterns of the fundamental rogue wave in the Sasa-Satsuma equation are far more diverse than most other integrable systems. As shown in \cite{Feng2022ShiZhangWu}, these fundamental rogue waves can be classified into at least four distinctive types which are determined by the values of the free parameter $a = \mathrm{i} \kappa$. On the other hand, among these four types, the so-called twisted rogue wave (TRW) pair (see Fig. \ref{fig:twisted RW}), which was first reported in \cite{Chen2013}, is a distinctive characteristic of the Sasa-Satsuma equation in contrast with many other integrable equations. As illustrated in Fig. \ref{fig:twisted RW}, this TRW pair possesses four zero-amplitude points $P_1, P_2, Q_1, Q_2 $ and two maximum amplitude points $M_1$ and $M_2$.
The appearance demonstrates that it comprises two extended rogue-wave components bending towards each other and displaying an identical but antisymmetric structure, which is the reason to name it as TRW pair. As depicted in \cite{Wu2022WeiShiFeng}, the fundamental rogue waves have three other intrinsic structures (see Fig. \ref{fig:first-order RW}) which can be obtained by altering the parameter values that appear in the solutions.

\begin{figure}[h]
    \centering
\subfigure[]{%
        \includegraphics[width=65mm]{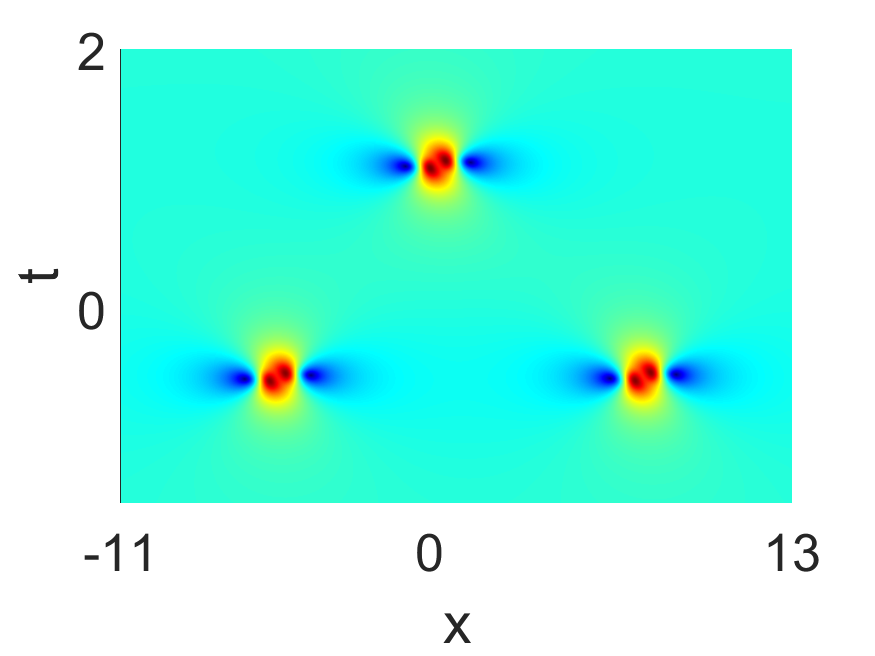}
        \label{fig:figure1(a)}}
\subfigure[]{%
        \includegraphics[width=65mm]{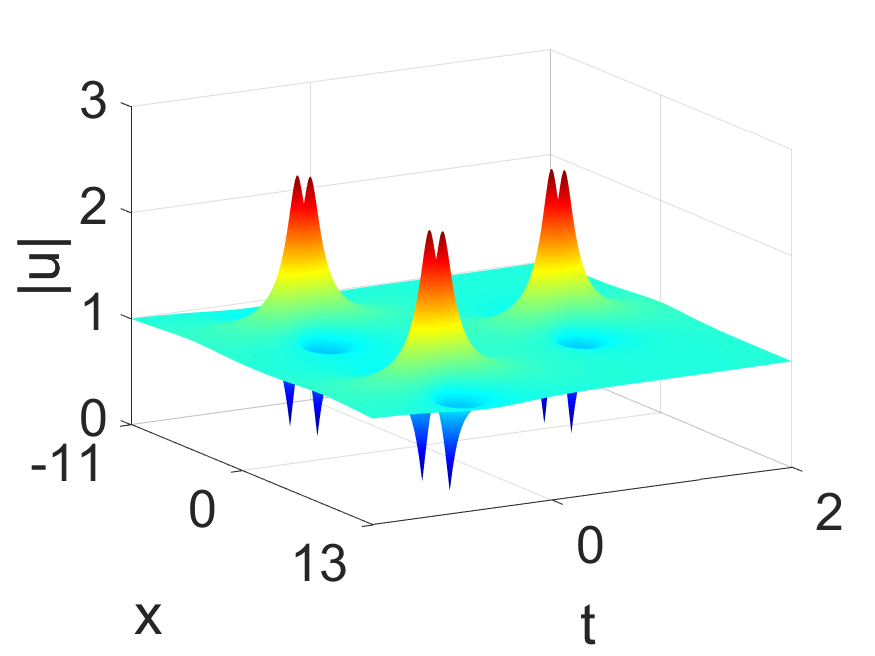}
        \label{fig:figure1(b)}}
    \caption{(Color online) Second-order rouge wave with parameter values \(a=1.4\mathrm{i},c=-0.7,a_1=a_2=0,a_3=100\exp(7\pi\mathrm{i}/4)\).}
    \label{fig:2nd RW-separate}
\end{figure}

\begin{figure}[h]
    \centering
\subfigure[]{%
        \includegraphics[width=65mm]{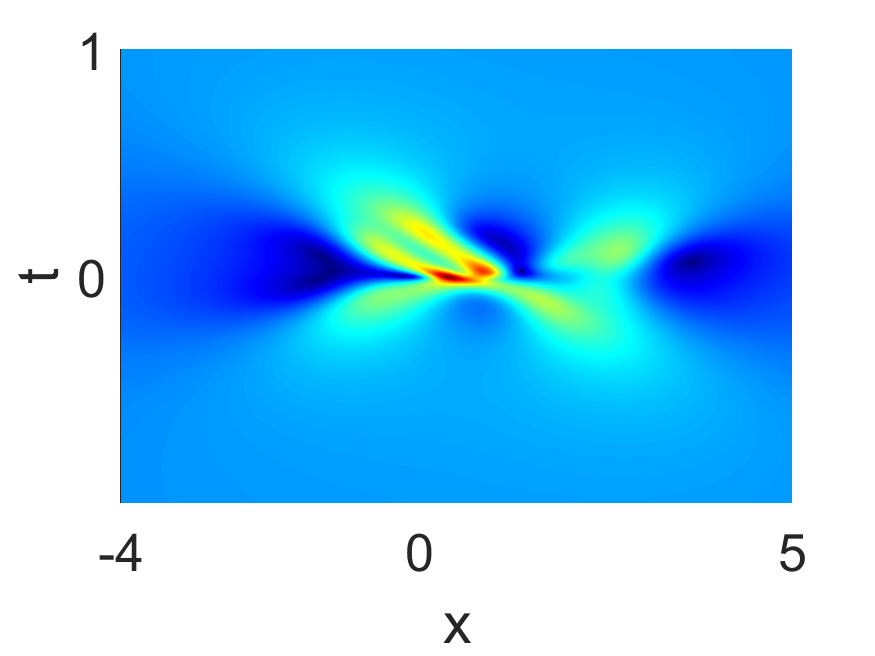}
        \label{fig:figure2(a)}}
\subfigure[]{%
        \includegraphics[width=65mm]{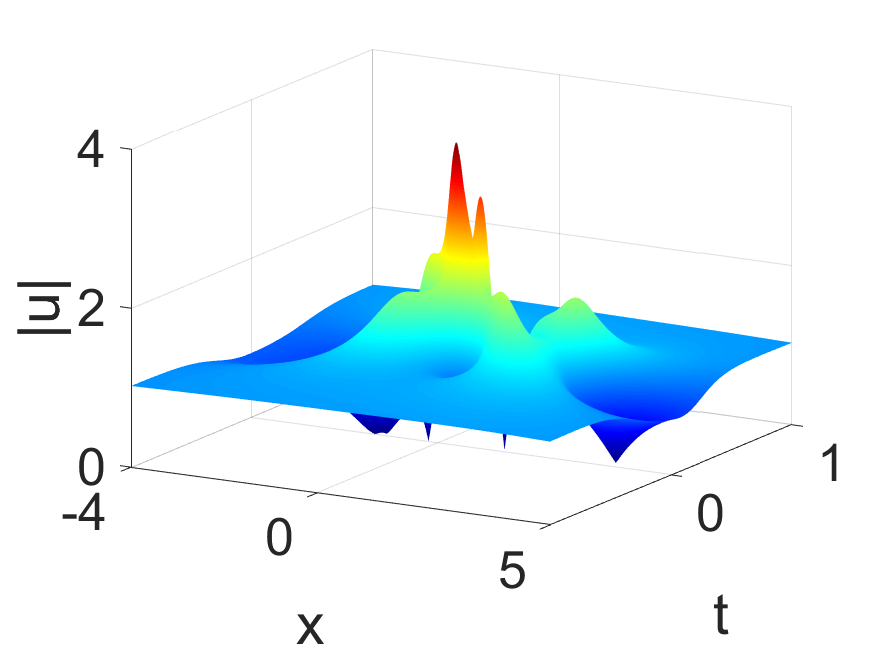}
        \label{fig:figure2(b)}}
    \caption{(Color online) Second-order rouge wave with parameter values \(a=1.4\mathrm{i},c=-0.7,a_1=a_2=3=0\).}
    \label{fig:2nd RW-collison}
\end{figure}

To get higher-order rogue waves, we may take $N\geq2$ in Theorem \ref{thm-Schur polynomials}. In particular, when $N=2$, the $\tau$-function is the determinant of a $4 \times 4$ matrix. Similar to the fundamental rogue waves, after some computations, we can show that the second-order rogue waves presented in Theorem \ref{thm-Schur polynomials} are equivalent to those given in Theorem \ref{thm-recursive relation}. As the solution dynamics of the second- and third-order rogue waves have been thoroughly investigated in \cite{Feng2022ShiZhangWu}, we only provide the graphs of second-order rogue waves in Figs \ref{fig:2nd RW-separate}-\ref{fig:2nd RW-collison}.

\begin{figure}[h]
    \centering
\subfigure[]{%
        \includegraphics[width=75mm]{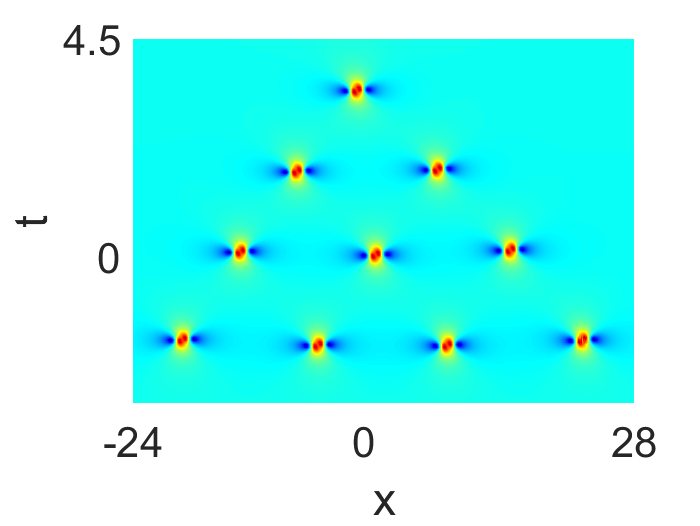}}
\subfigure[]{%
        \includegraphics[width=75mm]{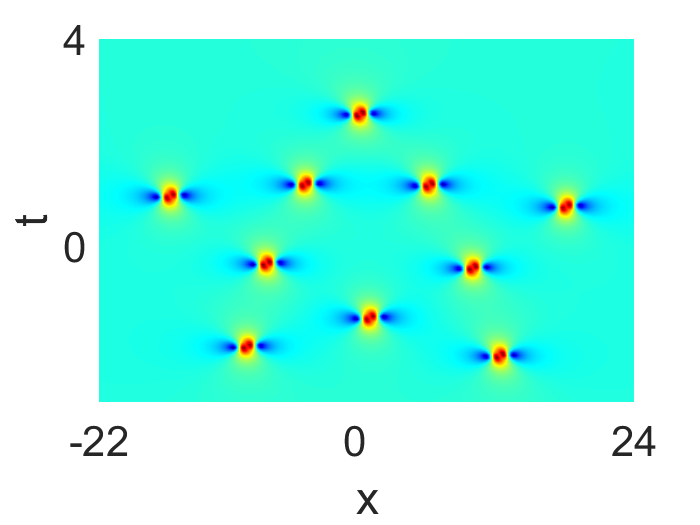}}
\subfigure[]{%
        \includegraphics[width=75mm]{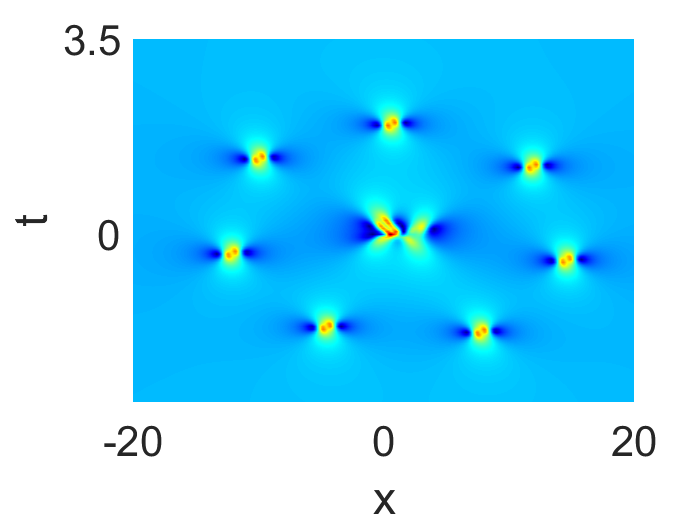}}
    \caption{(Color online) Fourth-order rouge waves under parameter values \(a = 1.4\mathrm{i}, c = -0.7, a_1=a_2=a_4=a_6= 0\) (a) \(a_3 = 200 \exp{(0.75\pi \mathrm{i})}, a_5 = 0, a_7 = 0\), (b) \(a_3 = 0, a_5 = 3000 \exp{(0.85\pi \mathrm{i})}, a_7 = 0\) and (c) \(a_3 = 0, a_5 = 0, a_7 = - 25000 \). }
    \label{fig:4RW}
\end{figure}

\begin{figure}[htp]
	\centering  
	\subfigbottomskip=2pt 
	\subfigcapskip=-5pt 
	\subfigure[ ]{
		\includegraphics[height=60mm,width=75mm]{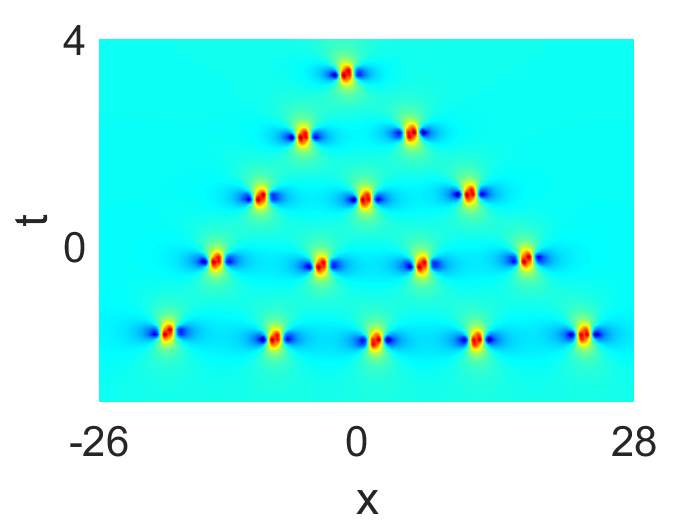}}
	\subfigure[ ]{
		\includegraphics[height=60mm,width=75mm]{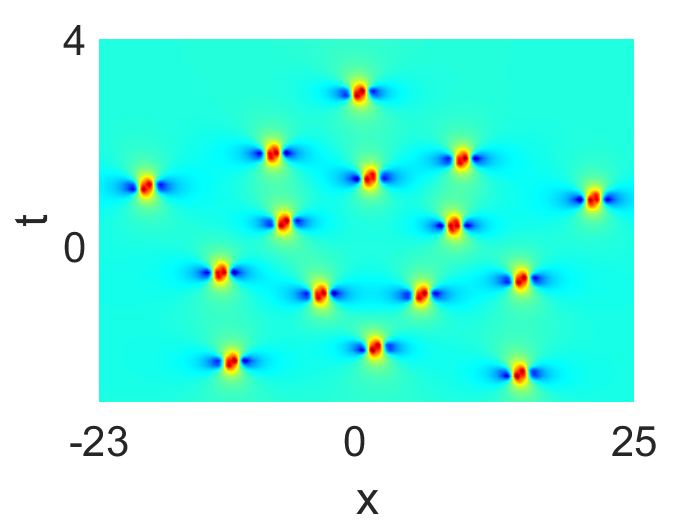}}
	  \\
	\subfigure[ ]{
		\includegraphics[height=60mm,width=75mm]{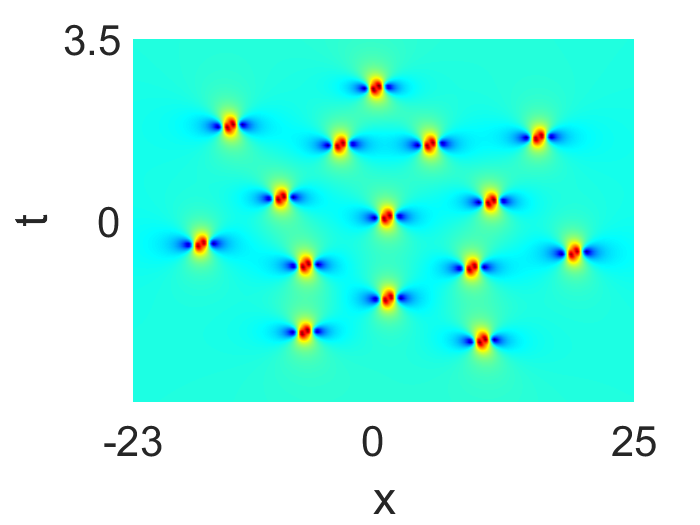}}
	\subfigure[ ]{
		\includegraphics[height=60mm,width=75mm]{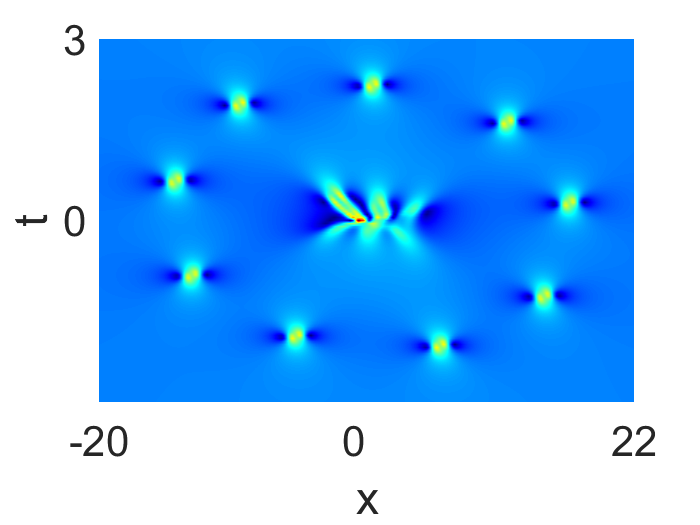}}
	\caption{(Color online) Fifth-order rouge waves under parameter values \(a = 1.4\mathrm{i}, c = -0.7, a_1=a_2=a_4=a_6=a_8= 0\) (a) \(a_3 = 100 \exp{(0.75\pi \mathrm{i})}, a_5 = 0, a_7 = 0, a_9=0\), (b) \(a_3 = 0, a_5 = 1500 \exp{(0.85\pi \mathrm{i})}, a_7 = 0,a_9=0\) , (c) \(a_3 = 0, a_5 = 0, a_7 = - 20000 ,a_9=0\) and (d) \(a_3 = 0, a_5 = 0, a_7 = 0 ,a_9=-200000\).  }
	\label{fig:5RW}
\end{figure}

\begin{figure}[htp]
\centering
\renewcommand\arraystretch{1.25}
\setlength\tabcolsep{1.25pt}
\begin{tabular}{m{0.5cm}<{\centering}m{3.8cm}<{\centering}m{3.8cm}<{\centering}m{3.8cm}<{\centering}m{3.8cm}<{\centering}r} &\textbf{Large $a_3$}          &    &  & &\\
\rotatebox{90}{\textbf{$N=2$}}&\includegraphics[height=32mm,width=36mm]{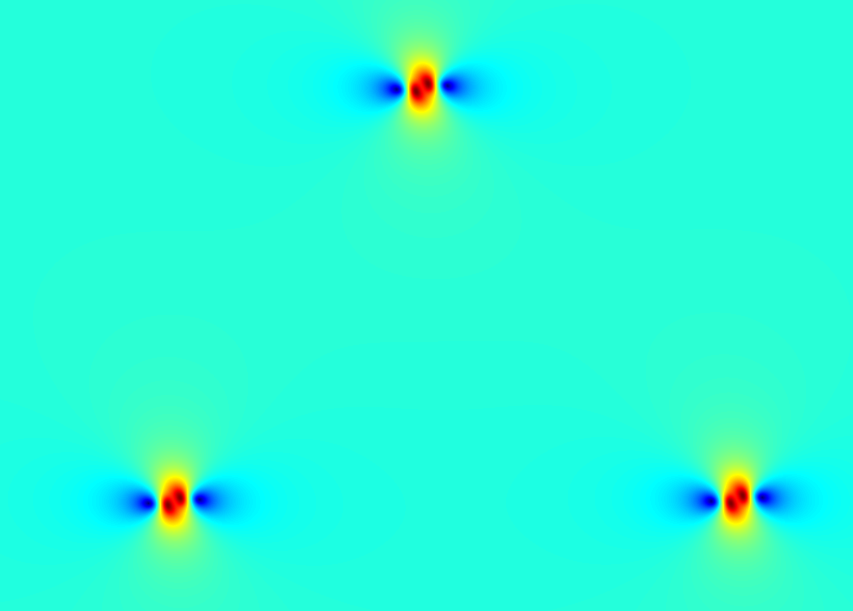}&\makecell{\\ \\ \\ \textbf{Large $a_5$}}\\
\rotatebox{90}{\textbf{$N=3$}}&\includegraphics[height=32mm,width=36mm]{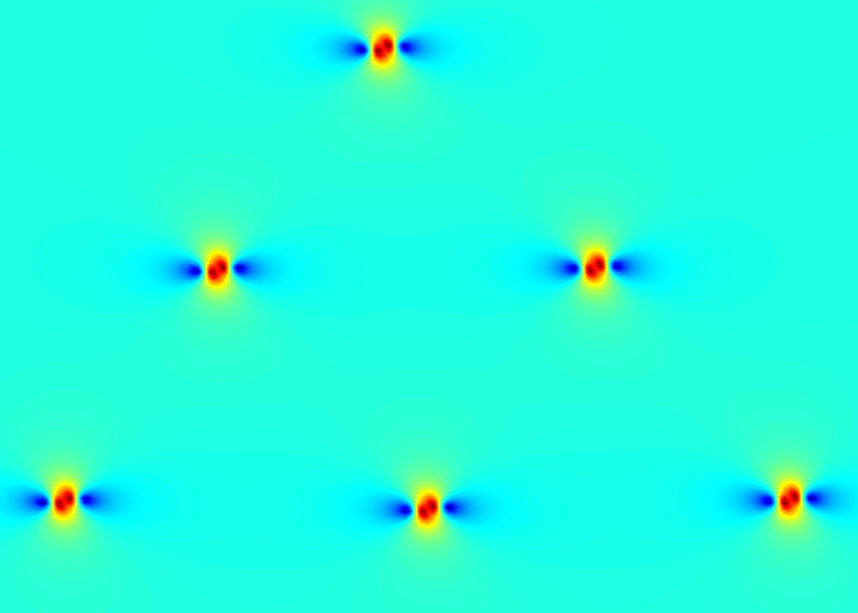}&\includegraphics[height=32mm,width=36mm]{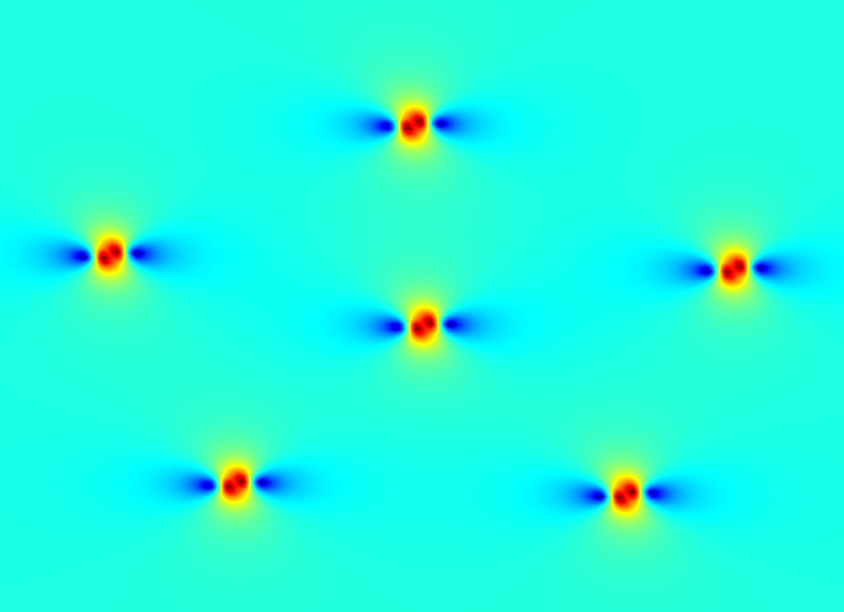}&\makecell{\\ \\ \\ \textbf{Large $a_7$}}&   &\\\rotatebox{90}{\textbf{$N=4$}}&\includegraphics[height=32mm,width=36mm]{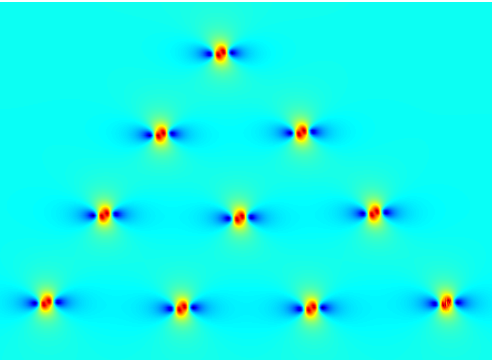}&\includegraphics[height=32mm,width=36mm]{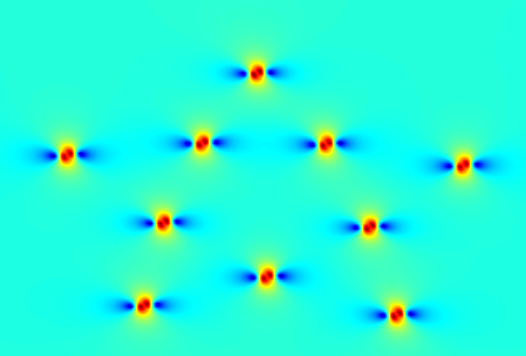}&\includegraphics[height=32mm,width=36mm]{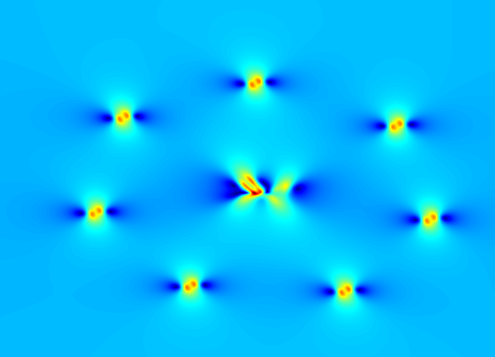}&\makecell{\\ \\ \\ \textbf{Large $a_9$}}&\\
\rotatebox{90}{\textbf{$N=5$}}&\includegraphics[height=32mm,width=36mm]{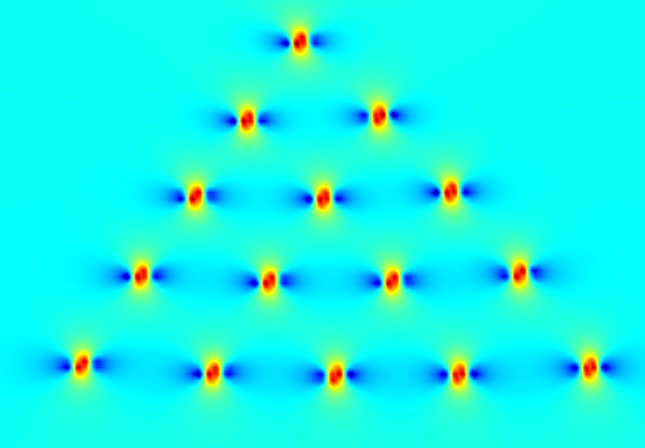}&\includegraphics[height=32mm,width=36mm]{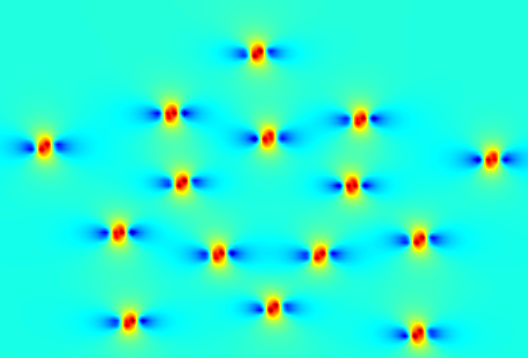}&\includegraphics[height=32mm,width=36mm]{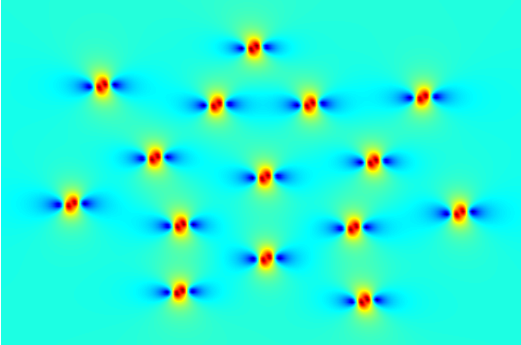}&\includegraphics[height=32mm,width=36mm]{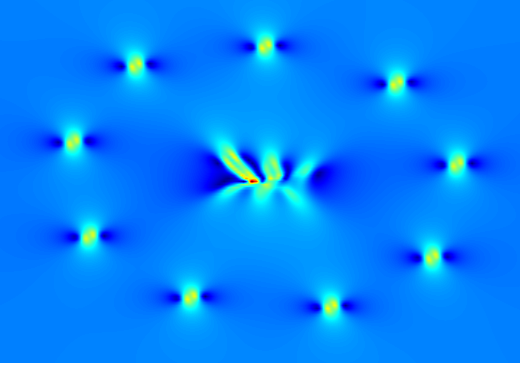}&$t$\\
 &$x$&$x$&$x$&$x$
\end{tabular}
\begin{tabular}{ccccc}
\hline$N$ & $a_{3}$ & $a_{5}$ & $a_{7}$ & $a_{9}$ \\
\hline 2 & $800\exp{(0.75\pi \mathrm{i})}$ & & & \\
3 & $500\exp{(0.75\pi \mathrm{i})}$ & $8000\exp{(0.85\pi \mathrm{i})}$ & & \\
4 & $200\exp{(0.75\pi \mathrm{i})}$ & $3000\exp{(0.85\pi \mathrm{i})}$ & $-25000$ & \\
5 & $100\exp{(0.75\pi \mathrm{i})}$ & $1500\exp{(0.85\pi \mathrm{i})}$ & $-20000$ & $-200000$\\
\hline
\end{tabular}

\caption{Predicted rogue patterns $u_N (x, t)$ for the orders $2 \le N \le5$ and the large parameter $a_{2m+1}$ from $m = 1$ to $N-1$. For each panel, the large parameter
$a_{2m+1}$ in the rogue wave is displayed in the above table, with the other internal parameters set as zero.}
\newpage
\label{UP}
\end{figure}

Recently Yang and Yang \cite{Yang2021Yang-1,Yang2021Yang-2} have performed a comprehensive study on the rogue wave patterns of several integrable equations including the nonlinear Schr\"{o}dinger equation, the Bossinesq equation and the Manakov system. They have shown that when one of the internal parameters in the rogue wave solutions is large enough,  the rogue waves may display universal patterns. In particular, they have shown that the fourth-order rogue wave may consist of ten separating fundamental rogue waves which are far away from the origin otherwise it is comprised of a second-order order super rogue wave near the origin and seven separating fundamental rogue waves that are far away from the origin.  Similarly, the fifth-order rogue wave is either comprised of fifteen separating fundamental rogue waves that are far away from the origin or made up of a third-order order super rogue wave near the origin and nine separating fundamental rogue waves that are far away from the origin. It turns out that this also occurs in the fourth- and fifth-order rogue wave solutions of the SS equation. As there are four types of fundamental rogue waves, we have totally four types of fourth-order rogue waves for each pattern. For simplicity, we merely show one type in Figs. \ref{fig:4RW} and \ref{fig:5RW}.  In addition, the rogue wave pattern up to fifth-order of the SS equation is summarized in Fig. \ref{UP}.


\newpage
{\color{white} a}

\section{Proof of Theorem \ref{thm-recursive relation}} \label{Proof of Theorem-recursive relation}

In this section, we derive the rogue wave solutions presented in Theorem \ref{thm-recursive relation}. First,  we transform the Sasa-Satsuma equation \eqref{SS equation} into bilinear forms. Then rogue wave solutions will be derived by connecting these bilinear equations with those coming from the KP hierarchy.

\begin{proposition}[\cite{Wu2022WeiShiFeng}] The Sasa-Satsuma equation
\begin{equation*}
u_{t}=u_{x x x}-6 c|u|^{2} u_{x}-3 c u\left(|u|^{2}\right)_{x}
\end{equation*}
can be transformed into the system of bilinear equations
\begin{equation} \label{bilinear form-SS}
\begin{aligned}
    &\left(D_{x}^{2}-4c\right) f \cdot f=-4 c g g^{*}\\
    &\left(D_{x}^{3}-D_{t}+3 \mathrm{i} \kappa D_{x}^{2}-3\left(\kappa^{2}+ 4 c \right) D_{x} - 6 \mathrm{i} \kappa c \right) g \cdot f+6 \mathrm{i} \kappa c q g=0\\
    &\left(D_{x}+2 \mathrm{i} \kappa\right) g \cdot g^{*}=2 \mathrm{i} \kappa q f\\
    \end{aligned}
\end{equation}
via the variable transformation
\begin{equation} \label{transformation1}
    u=\frac{g}{f} e^{\mathrm{i}\left(\kappa(x-6 c t)-\kappa^{3} t\right)},
\end{equation}
where   $\kappa$ is real, $f$ is a real-valued function, $g$ is a complex-valued function, $q$ is an  auxiliary  function  and $D$ is the Hirota's bilinear operator \cite{Hirota2004} defined by
\begin{eqnarray*}\label{doperator}
D_x^mD_t^nf\cdot g=\left.\left(\frac{\partial}{\partial x}-\frac{\partial}{\partial {x'}}\right)^m\left(\frac{\partial}{\partial t}-\frac{\partial}{\partial{t'}}\right)^n
[f(x,t)g(x',t')]\right|_{x'=x,t'=t}.
\end{eqnarray*}
\end{proposition}

Suppose $ \widetilde{m}_{i j}^{kl}, \varphi_{i}^{k l}, \psi_{j}^{k l}$, which are functions of $x,y,t,r $ and $s$, satisfy the differential and difference relations
\begin{eqnarray}
&&\partial_x \widetilde{m}_{i j}^{kl}= \varphi_i^{kl}\psi_{j}^{k,l} , \label{differential and difference relations-1} \\
&&    \partial_x\varphi_i^{k,l} =   \varphi_i^{k+1,l}+ a \varphi_i^{k,l}=  \varphi_i^{k,l+1}+ b\varphi_i^{k,l},
\\
&&\partial_x\psi_{j}^{k,l} = - \psi_{j}^{k-1,l} - a\psi_{j}^{k,l} =-\psi_{j}^{k,l-1} - b\psi_{j}^{k,l}, \label{Applied condition 1}
\\
&&    \partial_y\varphi_i^{k,l} =  \partial_{x}^{2}\varphi_{i}^{k,l} ,  \quad
    \partial_y\psi_{j}^{k,l} =  -\partial_{x}^{2}\psi_{j}^{k,l} ,\\
 &&   \partial_t\varphi_i^{k,l} =  \partial_{x}^{3}\varphi_{i}^{k,l} , \quad
    \partial_t\psi_{j}^{k,l}= \partial_{x}^{3}\psi_{j}^{k,l} , \\
 &&    \partial_r\varphi_{i}^{k,l}= \varphi_i^{k-1,l}, \quad
    \partial_r\psi_{j}^{k,l} = -\psi_{j}^{k+1,l},\\
 &&    \partial_s\varphi_i^{k,l} = \varphi_i^{k,l-1}, \quad
    \partial_s\psi_{j}^{k,l} = -\psi_{j}^{k,l+1}, \label{differential and difference relations-last 1}
\end{eqnarray}
which imply the relations
\begin{eqnarray}
 \partial_y \widetilde{m}_{i j}^{kl} &=& 
\varphi_{i}^{k+1,l}\psi_{j}^{kl}+\varphi_{i}^{kl}\psi_{j}^{k-1,l}+2a\varphi_{i}^{kl}\psi_{j}^{kl}
=
\varphi_{i}^{k,l+1}\psi_{j}^{kl}+\varphi_{i}^{kl}\psi_{j}^{k,l-1}+2b\varphi_{i}^{kl}\psi_{j}^{kl},
\\
   \partial_t \widetilde{m}_{i j}^{kl} &=& 
\varphi_{i}^{k+2,l}\psi_{j}^{kl}+3a\varphi_{i}^{k+1,l}\psi_{j}^{k,l}+\varphi_{i}^{k+1,l}\psi_{j}^{k-1,l}+3a^{2}\varphi_{i}^{kl}\psi_{j}^{kl}+3a\varphi_{i}^{kl}\psi_{j}^{k-1,l}+\varphi_{i}^{kl}\psi_{j}^{k-2,l} \qquad
\\&=&
\varphi_{i}^{k,l+2}\psi_{j}^{kl}+3b\varphi_{i}^{k,l+1}\psi_{j}^{kl}+\varphi_{i}^{k,l+1}\psi_{j}^{k,l-1}+3b^{2}\varphi_{i}^{kl}\psi_{j}^{kl}+3b\varphi_{i}^{kl}\psi_{j}^{k,l-1}+\varphi_{i}^{kl}\psi_{j}^{k,l-2},
\\
  \partial_r \widetilde{m}_{i j}^{kl} &=& - \varphi_i^{k-1,l}\psi_{j}^{k+1,l}, \quad
\partial_s \widetilde{m}_{i j}^{kl} =  -\varphi_i^{k,l-1}\psi_{j}^{k,l+1}\\
    \widetilde{m}_{i j}^{k+1,l} &=&   \widetilde{m}_{i j}^{k l} +\varphi_i^{kl}\psi_{j}^{k+1,l}, \quad
     \widetilde{m}_{i j}^{k,l+1} =  \widetilde{m}_{i j}^{k l} +\varphi_i^{kl}\psi_{j}^{k,l+1}
\end{eqnarray}
where $a,b$ are constants, then it can be calculated that \cite{Ohta2012Yang,Yang2021Yang-3} the determinant
\begin{eqnarray*}
\widetilde{\tau}_{kl} &=& \det_{1 \leq i,j \leq 2N}\left( \widetilde{m}_{ij}^{kl} \right)
\end{eqnarray*}
would satisfy the bilinear equations 
\begin{eqnarray}
&&\left(D_{r} D_{x}-2\right) \tau_{k l} \cdot \tau_{k l}=-2 \tau_{k+1, l} \tau_{k-1, l} \label{KP-1} \\
&&\left(D_{s} D_{x}-2\right) \tau_{k l} \cdot \tau_{k l}=-2 \tau_{k, l+1} \tau_{k, l-1} \label{KP-2} \\
&&\left(D_{x}^{2}-D_{y}+2 a D_{x}\right) \tau_{k+1, l} \cdot \tau_{k l}=0 \label{KP-3}\\
&&\left(D_{x}^{2}-D_{y}+2 b D_{x}\right) \tau_{k, l+1} \cdot \tau_{k l}=0 \label{KP-4}\\
&&\left(D_{x}^{3}+3 D_{x} D_{y}-4 D_{t}+3 a\left(D_{x}^{2}+D_{y}\right)+6 a^{2} D_{x}\right) \tau_{k+1, l} \cdot \tau_{k l}=0 \label{KP-5}\\
&&\left(D_{x}^{3}+3 D_{x} D_{y}-4 D_{t}+3 b\left(D_{x}^{2}+D_{y}\right)+6 b^{2} D_{x}\right) \tau_{k, l+1} \cdot \tau_{k l}=0 \label{KP-6}\\
&&\left(D_{r}\left(D_{x}^{2}-D_{y}+2 a D_{x}\right)-4 D_{x}\right) \tau_{k+1, l} \cdot \tau_{k l}=0 \label{KP-7}\\
&&\left(D_{s}\left(D_{x}^{2}-D_{y}+2 b D_{x}\right)-4 D_{x}\right) \tau_{k, l+1} \cdot \tau_{k l}=0 \label{KP-8}\\
&&\left(D_{s}\left(D_{x}^{2}-D_{y}+2 a D_{x}\right)-4\left(D_{x}+a-b\right)\right) \tau_{k+1, l} \cdot \tau_{k l}+4(a-b) \tau_{k+1, l+1} \tau_{k, l-1}=0 {\color{white}aaaaa}  \label{KP-9}\\
&&\left(D_{r}\left(D_{x}^{2}-D_{y}+2 b D_{x}\right)-4\left(D_{x}+b-a\right)\right) \tau_{k, l+1} \cdot \tau_{k l}+4(b-a) \tau_{k+1, l+1} \tau_{k-1, l}=0\label{KP-10}\\
&&\left(D_{x}+a-b\right) \tau_{k+1, l} \cdot \tau_{k, l+1}=(a-b) \tau_{k+1, l+1} \tau_{k l} \label{KP-11}.
\end{eqnarray}
Denote by
\begin{eqnarray}
m_{\alpha \beta}^{k l}&=&\frac{1}{p_{\alpha}+q_{\beta}}\left(-\dfrac{p_{\alpha}-a}{q_{\beta}+a}\right)^{k} \left(-\dfrac{p_{\alpha}-b}{q_{\beta}+b}\right)^{l}  e^{\xi_{\alpha}+\eta_{\beta}} \label{matrix element-without differential operator-1} \\
\overline{\varphi}_{\alpha }^{k l}&=&\left(p_{\alpha}-a\right)^{k} (p_{\alpha}-b)^{l} e^{\xi_{\alpha}} \label{matrix element-without differential operator-2}  \\
\overline{\psi}_{ \beta}^{k l}&=&\left[-(q_{\beta}+a)\right]^{-k} \left[-(q_{\beta}+b)\right]^{-l} e^{\eta_{\beta}} \label{matrix element-without differential operator-3}  \\
\xi_{\alpha}&=&p_{\alpha} x+p_{\alpha}^{2} y+p_{\alpha}^{3} t+\frac{1}{p_{\alpha}-a} r+\frac{1}{p_{\alpha}-b} s+\xi_{\alpha 0} \\
\eta_{\beta}&=&q_{\beta} x-q_{\beta}^{2} y+q_{\beta}^{3} t+\frac{1}{q_{\beta}+a} r+\frac{1}{q_{\beta}+b} s+\eta_{\beta 0}
\end{eqnarray}
where $\alpha, \beta =1,2,$ and $ p_{\alpha}, q_{\beta}, \xi_{\alpha 0}, \eta_{\beta 0}$ are constants,
then direct computations indicate that $m_{\alpha \beta}^{k l}, \overline{\varphi}_{\alpha }^{k l}, \overline{\psi}_{ \beta}^{k l}$ satisfy the differential and difference relations
\eqref{differential and difference relations-1}-\eqref{differential and difference relations-last 1}.

Next, we define
\begin{eqnarray*}
M_{ij}^{(k,l,\mu_i,\nu_j)} &=& \left(
  \begin{array}{cc}
     \widetilde{m}_{2i-1,2j-1}^{(k,l,\mu_i,\nu_j)} & \widetilde{m}_{2i-1,2j}^{(k,l,\mu_i,\nu_j)} \\
    \widetilde{m}_{2i,2j-1}^{(k,l,\mu_i,\nu_j)} & \widetilde{m}_{2i,2j}^{(k,l,\mu_i,\nu_j)} \\
  \end{array}
\right)
\\
 &=& \left(
\begin{array}{cccc}
   A_{i}^{(\mu_i)}(p_1) B_{j}^{(\nu_j)}(q_1) m_{11}^{kl} &  A_{i}^{(\mu_i)}(p_1) B_{j}^{(\nu_j)}(q_2) m_{12}^{kl}   \\
 A_{i}^{(\mu_i)}(p_2) B_{j}^{(\nu_j)}(q_1) m_{21}^{kl} &  A_{i}^{(\mu_i)}(p_2) B_{j}^{(\nu_j)}(q_2) m_{22}^{kl}   \\
\end{array}
\right)
\end{eqnarray*}
and
\begin{eqnarray*}
\left(
  \begin{array}{c}
     \varphi_{2i-1}^{(k,l,\mu_i)} \\  \varphi^{(k,l,\mu_i)}_{2i}
  \end{array}
\right)
 &=& \left(
\begin{array}{c}
   A_{i}^{(\mu_i)}(p_1)  \overline{\varphi}_{1}^{kl} \\  A_{i}^{(\mu_i)}(p_2)  \overline{\varphi}_{2}^{kl}
\end{array}
\right)
,
\\
\left(
  \begin{array}{cc}
     \psi_{2j-1}^{(k,l,\nu_j)}, \psi^{(k,l,\nu_j)}_{2j}
  \end{array}
\right)
 &=&  \left(
\begin{array}{cc}
   B_{j}^{(\nu_j)}(q_1) \overline{\psi}_{1}^{kl}, B_{j}^{(\nu_j)}(q_2)  \overline{\psi}_{2}^{kl}
\end{array}
\right)
\end{eqnarray*}
where $i,j = 1,2\cdots, N$ and
\begin{eqnarray*}
A_{i}^{(\mu_i)}(p) &=& \sum_{n=0}^{i} \dfrac{a_{i-n}^{(\mu_i)}(p)}{n!}  (p\partial_p)^{n}, \quad B_{j}^{(\nu_j)}(q) = \sum_{n=0}^{j} \dfrac{b_{j-n}^{(\nu_j)}(q)}{n!}  (q\partial_q)^{n}
\end{eqnarray*}
then the functions $ \widetilde{m}_{i j}^{(k,l,\mu_i)}, \varphi_{i}^{(k,l,\mu_j)}, \psi_{j}^{(k,l,\nu_j)}$ would satisfy the differential and difference relations
\eqref{differential and difference relations-1}-\eqref{differential and difference relations-last 1} as the operators $A_{i}^{(\mu_i)}$ and $B_{j}^{(\nu_j)}$ commute with the partial differentials with respect to $x,y,t,r,s$. Hence, the determinant
\begin{eqnarray*}
\widetilde{\tau}_{kl} &=&   \det_{1\leq \kappa,\lambda \leq N} \left(M_{i_{\kappa}, j_{\lambda}}^{(k,l,\mu_i,\nu_j)} \right)  
\end{eqnarray*}
satisfies the bilinear equations \eqref{KP-1}-\eqref{KP-11} for any sequence of indices $\left(i_{1}, i_{2}, \ldots, i_{N} ; j_{1}, j_{2}, \ldots, j_{N}\right)$. In particular, the determinant
\begin{eqnarray*}
\tau_{kl} &=&   \det_{1\leq i,j \leq N} \left(M_{2i-1, 2j-1}^{(k,l,\mu_i,\nu_j)} \right)  
\end{eqnarray*}
solves the bllinear equations \eqref{KP-1}-\eqref{KP-11}. 

From now on, for some reason that will be clear later, we set $\mu_i = N-i, \nu_j = N-j$ for fixed $N$ and
\begin{eqnarray*}
a_k^{(\mu+1)}(p) &=& \sum_{n=0}^k \dfrac{(p\partial_p)^{n+2}F(p)}{(n+2)!}  a_{k-n}^{(\mu)}(p),
 \\
   b_k^{(\nu+1)}(p) &=& \sum_{n=0}^k \dfrac{ (q\partial_q)^{n+2}F(q)}{(n+2)!}  b_{k-n}^{(\nu)}(q), \quad \mu,\nu=0,1,\cdots, N-1. 
\end{eqnarray*}
For brevity, we denote $M_{ij}^{(k,l,\mu_i,\nu_j)}$ and $ \widetilde{m}_{ij}^{(k,l,\mu_i,\nu_j)}$ by $M_{ij}^{kl}$ and $\widetilde{m}_{ij}^{kl}$ respectively.

In what follows, we will establish the reductions from the bilinear equations \eqref{KP-1}-\eqref{KP-11}
in the KP hierarchy to the bilinear equations \eqref{bilinear form-SS}. The reduction procedure consists of several steps. We start with the reduction from AKP to CKP \cite{Jimbo1983Miwa}. To achieve this, we impose the parameter relations
$$
q_{j}=p_{j},    \quad b=-a, \quad \xi_{j 0} = \eta_{j 0}, \quad a_n^{(0)} (p_j) = b_n^{(0)} (q_j),
$$
where $j= 1,2$ and $n=0,1,\dots,2N-1$, then we obtain
\begin{equation*}
   \xi_{j}(x, y, t, r, s)= \eta_{j} (x,-y, t, s, r)
\end{equation*}
and
\begin{equation*}
  a_k^{(\mu+1)}(p_{j}) = b_k^{(\mu+1)}(q_{j}).
\end{equation*}
Therefore, we have
\begin{equation*}
  m_{\beta \alpha }^{-l,-k} (x,-y, t, s, r) =  m_{\alpha \beta  }^{kl} (x, y, t, r, s), \quad 1 \leq \alpha, \beta \leq 2
\end{equation*}
and
\begin{eqnarray*}
&& M_{ji}^{-l,-k}(x,-y, t, s, r) \\
&=& \left(
\begin{array}{cccc}
   A_{2j-1}^{(N-j)}(p_1) B_{2i-1}^{(N-i)}(q_1) m_{11}^{-l,-k} &  A_{2j-1}^{(N-j)}(p_1) B_{2i-1}^{(N-i)}(q_2) m_{12}^{-l,-k}   \\
 A_{2j-1}^{(N-j)}(p_2) B_{2i-1}^{(N-i)}(q_1) m_{21}^{-l,-k} &  A_{2j-1}^{(N-j)}(p_2) B_{2i-1}^{(N-i)}(q_2) m_{22}^{-l,-k}   \\
\end{array}
\right)
(x,-y, t, s, r)
\\
&=& \left(
\begin{array}{cccc}
   B_{2j-1}^{(N-j)}(q_1) A_{2i-1}^{(N-i)}(p_1) m_{11}^{kl} &  B_{2j-1}^{(N-j)}(q_1) A_{2i-1}^{(N-i)}(p_2) m_{21}^{kl}   \\
 B_{2j-1}^{(N-j)}(q_2) A_{2i-1}^{(N-i)}(p_1) m_{12}^{kl} &  B_{2j-1}^{(N-j)}(q_2) A_{2i-1}^{(N-i)}(p_2) m_{22}^{kl}   \\
\end{array}
\right)
(x, y, t, r, s)
\\
&=& (M_{ij}^{kl})^T(x,y, t, r, s),
\end{eqnarray*}
where $M^T$ denotes the transpose of the matrix $M$.
Then it follows that
\begin{eqnarray}
\tau_{-l, -k}(x,-y, t, s, r) &=& |M_{ij}^{-l,-k}(x,-y, t, s, r)| \nonumber
\\
&=& |(M_{ji}^{kl})^T(x,y, t, r, s)|  \nonumber
\\
&=& |(\widetilde{m}_{ij}^{kl})^T|  \nonumber
\\
&=&
 \tau_{k l}(x, y, t, r, s). \label{symmetry}
\end{eqnarray}

Now we consider the dimension reduction that is a crucial step in the derivation of rogue wave solutions. We first introduce the operator defined by
\begin{equation}\label{differential operator-dimension reduction}
\widetilde{D}_{r,s,x} = \left(\partial_r + \partial_s - \dfrac{1}{c} \partial_x \right).
\end{equation}
Let
\begin{eqnarray*}
  F(p) &=& \dfrac{1}{p-a} + \dfrac{1}{p+a} - \dfrac{p}{c},
  \\
  G(p) &=& p\partial_p F(p) = - p \left[ \dfrac{1}{(p-a)^2} + \dfrac{1}{(p+a)^2} + \dfrac{1}{c} \right],
\end{eqnarray*}
then we have
\begin{eqnarray}
  B_1^{(N-1)} \widetilde{D}_{r,s,x}m &=& B_1^{(N-1)} [F(p) + F(q)] m  \nonumber \\
       &=& [F(p) + F(q)] B_1^{(N-1)} m +   b_0^{(N-1)} G(q) m \label{dimension reduction-D-b1-recursive relation}
\end{eqnarray}
and for $2 \leq j\leq N$,
\begin{eqnarray}
  B_{2j-1}^{(\nu)}(q) \widetilde{D}_{r,s,x}m &=& B_{2j-1}^{(\nu)}(q) [F(p) + F(q)] m  \nonumber \\
  &=&  \sum_{n=0}^{2j-1} \dfrac{b_{2j-1-n}^{(\nu)}(q)}{n!}  (q\partial_q)^{n}  [F(p) + F(q)] m \nonumber\\
    &=&  \sum_{n=0}^{2j-1} \dfrac{b_{2j-1-n}^{(\nu)}(q)}{n!} \sum_{k=0}^n           \left(
                                                                                   \begin{array}{c}
                                                                                     n \\
                                                                                     k \\
                                                                                   \end{array}
                                                                                      \right)  \{(q\partial_q)^{n-k}  [F(p) + F(q)]\} (q\partial_q)^k m \nonumber
                                                                                       \\
  &=& \sum_{n=0}^{2j-1} \dfrac{b_{2j-1-n}^{(\nu)}(q)}{n!}  [F(p) + F(q)] (q\partial_q)^n m  \nonumber \\
  &&+
   \sum_{n=1}^{2j-1} \dfrac{b_{2j-1-n}^{(\nu)}(q)}{(n-1)!}     [(q\partial_q)F(q)]       (q\partial_q)^{n-1} m \nonumber\\
&&+  \sum_{n=2}^{2j-1} \dfrac{b_{2j-1-n}^{(\nu)}(q)}{n!} \sum_{k=0}^{n-2}           \left(
                                                                                   \begin{array}{c}
                                                                                     n \\
                                                                                     k \\
                                                                                   \end{array}
                                                                                      \right)  \{(q\partial_q)^{n-k}  [F(p) + F(q)]\} (q\partial_q)^k m \nonumber\\
  &=& [F(p) + F(q)]  B_{2j-1}^{(\nu)}(q)  m +
  [(q\partial_q)F(q)] \sum_{n=1}^{2j-1} \dfrac{b_{2j-1-n}^{(\nu)}(q)}{(n-1)!}     (q\partial_q)^{n-1} m \nonumber\\
&&+ \sum_{k=0}^{2j-3} \dfrac{1}{k!} \left[   \sum_{n=k+2}^{2j-1} \dfrac{b_{2j-1-n}^{(\nu)}(q)}{(n-k)!}                                                                                          (q\partial_q)^{n-k}   F(q) \right] (q\partial_q)^k m  \nonumber \\
 &=& [F(p) + F(q)]  B_{2j-1}^{(\nu)}(q)  m +
  [(q\partial_q)F(q)] \sum_{n=1}^{2j-1} \dfrac{b_{2j-1-n}^{(\nu)}(q)}{(n-1)!}     (q\partial_q)^{n-1} m \nonumber\\
&&+ \sum_{k=0}^{2j-3} \dfrac{1}{k!} \left[   \sum_{m=0}^{2j-3-k} \dfrac{(q\partial_q)^{m+2}   F(q) }{(m+2)!}   b^{(\nu)}_{2j-3-k-m}(q)                                                                                       \right] (q\partial_q)^k m \nonumber
\\
 &=& [F(p) + F(q)]  B_{2j-1}^{(\nu)}(q)  m +
  [(q\partial_q)F(q)] \sum_{n=1}^{2j-1} \dfrac{b_{2j-1-n}^{(\nu)}(q)}{(n-1)!}     (q\partial_q)^{n-1} m \nonumber\\
&&+ \left[ \sum_{k=0}^{2j-3} \dfrac{1}{k!} b^{(\nu+1)}_{2j-3-k}(q)  (q\partial_q)^k \right] m \nonumber
\\
 &=& [F(p) + F(q)]  B_{2j-1}^{(\nu)}(q)  m +
  G(q) \sum_{n=1}^{2j-1} \dfrac{b_{2j-1-n}^{(\nu)}(q)}{(n-1)!}     (q\partial_q)^{n-1} m \nonumber\\
&&+ B_{2j-3}^{(\nu+1)}(q)  m.  \label{dimension reduction-D-bj-recursive relation}
\end{eqnarray}
Similarly, we can get
\begin{eqnarray}
A_1^{(N-1)} \widetilde{D}_{r,s,x}m &=& [F(p) + F(q)] A_1^{(N-1)} m +   a_0^{(N-1)} G(p) m \label{dimension reduction-D-a1-recursive relation}
\\
 A_{2i-1}^{(\mu)}(p) \widetilde{D}_{r,s,x}m  &=& [F(p) + F(q)]  A_{2i-1}^{(\mu)}(p)  m +
  G(p) \sum_{n=1}^{2i-1} \dfrac{a_{2i-1-n}^{(\mu)}(p)}{(n-1)!}     (p\partial_p)^{n-1} m \nonumber\\
&&+ A_{2i-3}^{(\mu+1)}(p)  m, \label{dimension reduction-D-ai-recursive relation}
\end{eqnarray}
where $2 \leq i\leq N$.

Since $\xi $ and $ \xi^*$ are a pair of complex conjugate roots of $F'(p) = 0$, we have
\begin{equation} \label{root of G}
G(\xi) = G(\xi^*) = 0.
\end{equation}
Then combing the equations \eqref{dimension reduction-D-b1-recursive relation}-\eqref{root of G} yields that
\begin{eqnarray*}
 \widetilde{D}_{r,s,x}A_{1}^{(\mu)}(p) B_{1}^{(\nu)}(q) m  \Big|_{\substack{p_1= q_1=\xi, \\ p_2=q_2= \xi^*}}
&=& [F(p) + F(q)]   A_{1}^{(\mu)}(p) B_{1}^{(\nu)}(q) m\Big|_{\substack{p_1= q_1=\xi, \\ p_2=q_2= \xi^*}}
\\
 \widetilde{D}_{r,s,x}A_{1}^{(\mu)}(p) B_{2j-1}^{(\nu)}(q) m\Big|_{\substack{p_1= q_1=\xi, \\ p_2=q_2= \xi^*}}
&=& [F(p) + F(q)]   A_{1}^{(\mu)}(p) B_{2j-1}^{(\nu)}(q) m\Big|_{\substack{p_1= q_1=\xi, \\ p_2=q_2= \xi^*}}
\\
&&+ A_{1}^{(\mu)}(p)   B_{2j-3}^{(\nu+1)}(q)  m\Big|_{\substack{p_1= q_1=\xi, \\ p_2=q_2= \xi^*}}   , \quad 2 \leq j\leq N
\\
 \widetilde{D}_{r,s,x}A_{2i-1}^{(\mu)}(p) B_{1}^{(\nu)}(q) m\Big|_{\substack{p_1= q_1=\xi, \\ p_2=q_2= \xi^*}}
&=& [F(p) + F(q)]   A_{2i-1}^{(\mu)}(p) B_{1}^{(\nu)}(q) m\Big|_{\substack{p_1= q_1=\xi, \\ p_2=q_2= \xi^*}}
\\
&&+ A_{2i-3}^{(\mu+1)}(p)   B_{1}^{(\nu)}(q)  m\Big|_{\substack{p_1= q_1=\xi, \\ p_2=q_2= \xi^*}}   , \quad 2 \leq i\leq N
\\
 \widetilde{D}_{r,s,x}A_{2i-1}^{(\mu)}(p) B_{2j-1}^{(\nu)}(q) m\Big|_{\substack{p_1= q_1=\xi, \\ p_2=q_2= \xi^*}}
    &=& [F(p) + F(q)]   A_{2i-1}^{(\mu)}(p)   B_{2j-1}^{(\nu)}(q)  m  + A_{2i-3}^{(\mu+1)}(p)   B_{2j-1}^{(\nu)}(q)  m
    \\
    && + A_{2i-1}^{(\mu)}(p)   B_{2j-3}^{(\nu+1)}(q)  m\Big|_{\substack{p_1= q_1=\xi, \\ p_2=q_2= \xi^*}}  , \quad 2 \leq i,j\leq N.
\end{eqnarray*}

As a consequence, if we define
\begin{equation*}
  \widetilde{m}_{-1,j}^{kl} = \widetilde{m}_{i,-1}^{kl} = 0, \quad i,j = 1,2,\dots,2N,
\end{equation*}
then it follows that
\begin{eqnarray}
  && \widetilde{D}_{r,s,x} \, \tau_{kl}\Big|_{\substack{p_1= q_1=\xi, \\ p_2=q_2= \xi^*}}
  \\
    &=& \sum^{2N}_{i,j=1}\Delta_{ij} ( \widetilde{D}_{r,s,x} \widetilde{m}_{ij}^{kl} )\Big|_{\substack{p_1= q_1=\xi, \\ p_2=q_2= \xi^*}}   \\
       &=& \sum^{N}_{i,j=1}\Delta_{2i-1,2j-1} \left( \left[F(p_1) + F(q_1)\right] \widetilde{m}_{2i-1,2j-1}^{kl} + \widetilde{m}_{2i-3,2j-1}^{kl}+\widetilde{m}_{2i-1,2j-3}^{kl} \right) \\
       &&+ \sum^{N}_{i,j=1}\Delta_{2i-1,2j} \left( \left[F(p_1) + F(q_2)\right] \widetilde{m}_{2i-1,2j}^{kl} + \widetilde{m}_{2i-3,2j}^{kl}+\widetilde{m}_{2i-1,2j-2}^{kl} \right) \\
       &&+ \sum^{N}_{i,j=1}\Delta_{2i,2j-1} \left( \left[F(p_2) + F(q_1)\right] \widetilde{m}_{2i,2j-1}^{kl} + \widetilde{m}_{2i-2,2j-1}^{kl}+\widetilde{m}_{2i,2j-3}^{kl} \right) \\
       &&+ \sum^{N}_{i,j=1}\Delta_{2i,2j} \left( \left[F(p_2) + F(q_2)\right] \widetilde{m}_{2i,2j}^{kl} + \widetilde{m}_{2i-2,2j-2}^{kl}+\widetilde{m}_{2i-2,2j-2}^{kl} \right)\Big|_{\substack{p_1= q_1=\xi, \\ p_2=q_2= \xi^*}}   \\
      &=&  F(p_1) \sum_{i=1}^N \sum_{j=1}^{2N} \Delta_{2i-1,j} \widetilde{m}_{2i-1,j}^{kl} + F(p_2) \sum_{i=1}^N \sum_{j=1}^{2N} \Delta_{2i,j} \widetilde{m}_{2i,j}^{kl}  \\
       &&+  F(q_1) \sum_{i=1}^{2N} \sum_{j=1}^{N} \Delta_{i,2j-1} \widetilde{m}_{i,2j-1}^{kl} + F(q_2) \sum_{i=1}^{2N} \sum_{j=1}^{N} \Delta_{i,2j} \widetilde{m}_{i,2j}^{kl}  \\
       &&+\sum^{N}_{i,j=1}\Delta_{2i-1,2j-1} \left( \widetilde{m}_{2i-3,2j-1}^{kl}+\widetilde{m}_{2i-1,2j-3}^{kl} \right) \\
       &&+ \sum^{N}_{i,j=1}\Delta_{2i-1,2j} \left( \widetilde{m}_{2i-3,2j}^{kl}+\widetilde{m}_{2i-1,2j-2}^{kl} \right) \\
       &&+ \sum^{N}_{i,j=1}\Delta_{2i,2j-1} \left(\widetilde{m}_{2i-2,2j-1}^{kl}+\widetilde{m}_{2i,2j-3}^{kl} \right) \\
       &&+ \sum^{N}_{i,j=1}\Delta_{2i,2j} \left(  \widetilde{m}_{2i-2,2j-2}^{kl}+\widetilde{m}_{2i-2,2j-2}^{kl} \right)\Big|_{\substack{p_1= q_1=\xi, \\ p_2=q_2= \xi^*}}   \\
    &=& \left[F(p_1) + F(q_1)+F(p_2) + F(q_2)\right] N  \tau_{kl}\Big|_{\substack{p_1= q_1=\xi, \\ p_2=q_2= \xi^*}}
\end{eqnarray}

Thus, with the above result, we can replace the derivatives in $r$ and $s$ by derivatives in $x$ in the bilinear equations \eqref{KP-1}-\eqref{KP-11}  and obtain
\begin{eqnarray}
&& \left(D_{x}^{2}-4 c\right) \tau_{k l} \cdot \tau_{k l}=-2 c\left(\tau_{k+1, l} \tau_{k-1, l}+\tau_{k, l+1} \tau_{k, l-1}\right) \label{reduced bilinear eq1}\\
&& \left(D_{x}^{3}-D_{t}+3 a D_{x}^{2}+3\left(a^{2}-2 c\right) D_{x}-6 a c\right) \tau_{k+1, l} \cdot \tau_{k l}+6 a c \tau_{k+1, l+1} \tau_{k, l-1}=0 \label{reduced bilinear eq2}\\
&& \left(D_{x}^{3}-D_{t}-3 a D_{x}^{2}+3\left(a^{2}-2 c\right) D_{x}+6 a c\right) \tau_{k, l+1} \cdot \tau_{k l}-6 a c \tau_{k+1, l+1} \tau_{k-1, l}=0 \label{reduced bilinear eq3}\\
&& \left(D_{x}+2 a\right) \tau_{k+1, l} \cdot \tau_{k, l+1}=2 a \tau_{k+1, l+1} \tau_{k l}. \label{reduced bilinear eq4}
\end{eqnarray}
Since the bilinear equations \eqref{reduced bilinear eq1}-\eqref{reduced bilinear eq4} do not involve derivatives with respect to $y,r$ and $s$, we may take $y=r=s=0$.
Then according to \eqref{symmetry}, we have
\begin{equation} \label{symmetry2}
\tau_{k l}(x, t)=\tau_{-l, -k}(x, t).
\end{equation}

Finally, we consider the complex conjugate reduction. 
Note that the parameter relations \eqref{parameter relation-recursive relation} imply that
\begin{eqnarray*}  \label{parameter relation-1-recursive relation}
   a_k^{(\mu)} (p_1) = [  a_k^{(\mu)} (p_2)]^*, \quad b_k^{(\nu)} (q_1) = [  b_k^{(\nu)} (p_2)]^*.
\end{eqnarray*}
Then using the parameter relations \eqref{parameter relation-recursive relation}, \eqref{parameter relation-1-recursive relation} and the assumption that $a=\mathrm{i} \kappa$  is purely imaginary, we may deduce that
 \begin{eqnarray*}
(M_{ij}^{0k})^* &=& \left(
\begin{array}{cccc}
   A_{2i-1}^{(N-i)}(p_1) B_{2j-1}^{(N-j)}(q_1) m_{11}^{0k} &  A_{2i-1}^{(N-i)}(p_1) B_{2j-1}^{(N-j)}(q_2) m_{12}^{0k}   \\
 A_{2i-1}^{(N-i)}(p_2) B_{2j-1}^{(N-j)}(q_1) m_{21}^{0k} &  A_{2i-1}^{(N-i)}(p_2) B_{2j-1}^{(N-j)}(q_2) m_{22}^{0k}   \\
\end{array}
\right)^*
\\&=& \left(
\begin{array}{cccc}
   A_{2i-1}^{(N-i)}(p_2) B_{2j-1}^{(N-j)}(q_2) m_{22}^{k0} &  A_{2i-1}^{(N-i)}(p_2) B_{2j-1}^{(N-j)}(q_1) m_{21}^{k0}   \\
 A_{2i-1}^{(N-i)}(p_1) B_{2j-1}^{(N-j)}(q_2) m_{12}^{k0} &  A_{2i-1}^{(N-i)}(p_1) B_{2j-1}^{(N-j)}(q_1) m_{11}^{k0}   \\
\end{array}
\right).
\end{eqnarray*}
As a consequence, we have
\begin{eqnarray*}
\tau_{0k}^* &=&  | \widetilde{m}_{ij}^{0k}|^*
\\
&=&  \left|
\begin{array}{cccccccc}
   \widetilde{m}_{22}^{k0} &\widetilde{m}_{21}^{k0} & \cdots & \widetilde{m}_{2,2j}^{k0} &\widetilde{m}_{2,2j-1}^{k0} & \cdots & \widetilde{m}_{2,2N}^{k0} &\widetilde{m}_{2,2N-1}^{k0}   \\
     \widetilde{m}_{12}^{k0} &\widetilde{m}_{11}^{k0} & \cdots & \widetilde{m}_{1,2j}^{k0} &\widetilde{m}_{1,2j-1}^{k0} & \cdots & \widetilde{m}_{1,2N}^{k0} &\widetilde{m}_{1,2N-1}^{k0}   \\
          \vdots &\vdots &  \ddots & \vdots &\vdots & \ddots   &\vdots & \vdots   \\
          \widetilde{m}_{2i,2}^{k0} &\widetilde{m}_{2i,1}^{k0} & \cdots & \widetilde{m}_{2i,2j}^{k0} &\widetilde{m}_{2i,2j-1}^{k0} & \cdots & \widetilde{m}_{2i,2N}^{k0} &\widetilde{m}_{2i,2N-1}^{k0}   \\
          \widetilde{m}_{2i-1,2}^{k0} &\widetilde{m}_{2i-1,1}^{k0} & \cdots & \widetilde{m}_{2i-1,2j}^{k0} &\widetilde{m}_{2i-1,2j-1}^{k0} & \cdots & \widetilde{m}_{2i-1,2N}^{k0} &\widetilde{m}_{2i-1,2N-1}^{k0}   \\
          \vdots &\vdots &  \ddots & \vdots &\vdots & \ddots   &\vdots & \vdots   \\
          \widetilde{m}_{2N,2}^{k0} &\widetilde{m}_{2N,1}^{k0} & \cdots & \widetilde{m}_{2N,2j}^{k0} &\widetilde{m}_{2N,2j-1}^{k0} & \cdots & \widetilde{m}_{2N,2N}^{k0} &\widetilde{m}_{2N,2N-1}^{k0}   \\
          \widetilde{m}_{2N-1,2}^{k0} &\widetilde{m}_{2N-1,1}^{k0} & \cdots & \widetilde{m}_{2N-1,2j}^{k0} &\widetilde{m}_{2N-1,2j-1}^{k0} & \cdots & \widetilde{m}_{2N-1,2N}^{k0} &\widetilde{m}_{2N-1,2N-1}^{k0}
\end{array}
\right|
\\
&=& (-1)^N \left| \begin{array}{cccccccc}
       \widetilde{m}_{12}^{k0} &\widetilde{m}_{11}^{k0} & \cdots & \widetilde{m}_{1,2j}^{k0} &\widetilde{m}_{1,2j-1}^{k0} & \cdots & \widetilde{m}_{1,2N}^{k0} &\widetilde{m}_{1,2N-1}^{k0}   \\
        \widetilde{m}_{22}^{k0} &\widetilde{m}_{21}^{k0} & \cdots & \widetilde{m}_{2,2j}^{k0} &\widetilde{m}_{2,2j-1}^{k0} & \cdots & \widetilde{m}_{2,2N}^{k0} &\widetilde{m}_{2,2N-1}^{k0}   \\
          \vdots &\vdots &  \ddots & \vdots &\vdots & \ddots   &\vdots & \vdots   \\
          \widetilde{m}_{2i-1,2}^{k0} &\widetilde{m}_{2i-1,1}^{k0} & \cdots & \widetilde{m}_{2i-1,2j}^{k0} &\widetilde{m}_{2i-1,2j-1}^{k0} & \cdots & \widetilde{m}_{2i-1,2N}^{k0} &\widetilde{m}_{2i-1,2N-1}^{k0}   \\
                    \widetilde{m}_{2i,2}^{k0} &\widetilde{m}_{2i,1}^{k0} & \cdots & \widetilde{m}_{2i,2j}^{k0} &\widetilde{m}_{2i,2j-1}^{k0} & \cdots & \widetilde{m}_{2i,2N}^{k0} &\widetilde{m}_{2i,2N-1}^{k0}   \\
          \vdots &\vdots &  \ddots & \vdots &\vdots & \ddots   &\vdots & \vdots   \\
          \widetilde{m}_{2N-1,2}^{k0} &\widetilde{m}_{2N-1,1}^{k0} & \cdots & \widetilde{m}_{2N-1,2j}^{k0} &\widetilde{m}_{2N-1,2j-1}^{k0} & \cdots & \widetilde{m}_{2N-1,2N}^{k0} &\widetilde{m}_{2N-1,2N-1}^{k0}\\
          \widetilde{m}_{2N,2}^{k0} &\widetilde{m}_{2N,1}^{k0} & \cdots & \widetilde{m}_{2N,2j}^{k0} &\widetilde{m}_{2N,2j-1}^{k0} & \cdots & \widetilde{m}_{2N,2N}^{k0} &\widetilde{m}_{2N,2N-1}^{k0}
\end{array}
\right|
\\
&=&  \left| \begin{array}{cccccccc}
       \widetilde{m}_{11}^{k0} & \widetilde{m}_{12}^{k0} &   \cdots &\widetilde{m}_{1,2j-1}^{k0} & \widetilde{m}_{1,2j}^{k0} & \cdots      &\widetilde{m}_{1,2N-1}^{k0}   & \widetilde{m}_{1,2N}^{k0} \\
        \widetilde{m}_{22}^{k0} &\widetilde{m}_{21}^{k0} & \cdots & \widetilde{m}_{2,2j}^{k0} &\widetilde{m}_{2,2j-1}^{k0} & \cdots  &\widetilde{m}_{2,2N-1}^{k0}  & \widetilde{m}_{2,2N}^{k0} \\
          \vdots &\vdots &  \ddots & \vdots &\vdots & \ddots   &\vdots & \vdots   \\
      \widetilde{m}_{2i-1,1}^{k0} &     \widetilde{m}_{2i-1,2}^{k0} &\cdots &\widetilde{m}_{2i-1,2j-1}^{k0} & \widetilde{m}_{2i-1,2j}^{k0}  & \cdots &\widetilde{m}_{2i-1,2N-1}^{k0} & \widetilde{m}_{2i-1,2N}^{k0}    \\
                    \widetilde{m}_{2i,1}^{k0} &\widetilde{m}_{2i,2}^{k0} &\cdots &\widetilde{m}_{2i,2j-1}^{k0} & \widetilde{m}_{2i,2j}^{k0} & \cdots &\widetilde{m}_{2i,2N-1}^{k0}& \widetilde{m}_{2i,2N}^{k0}   \\
          \vdots &\vdots &  \ddots & \vdots &\vdots & \ddots   &\vdots & \vdots   \\
          \widetilde{m}_{2N-1,1}^{k0} & \widetilde{m}_{2N-1,2}^{k0} &\cdots & \widetilde{m}_{2N-1,2j-1}^{k0} & \widetilde{m}_{2N-1,2j}^{k0} & \cdots &\widetilde{m}_{2N-1,2N-1}^{k0} & \widetilde{m}_{2N-1,2N}^{k0} \\
       \widetilde{m}_{2N,1}^{k0} &   \widetilde{m}_{2N,2}^{k0} & \cdots & \widetilde{m}_{2N,2j-1}^{k0} & \widetilde{m}_{2N,2j}^{k0} &  \cdots &\widetilde{m}_{2N,2N-1}^{k0} & \widetilde{m}_{2N,2N}^{k0}
\end{array}
\right|
\\
&=& \tau_{k0}.
\end{eqnarray*}
Similar method as above demonstrates that
$$
\tau^*_{k k } =  \tau_{ k k},
$$
which implies that $\tau_{ k k}$ is real.
Define
$$
\widetilde{f}=\tau_{00}, \quad \widetilde{g}=\tau_{10}, \quad \widetilde{h}=\tau_{01}, \quad \widetilde{q}=\tau_{11},
$$
then it is clear that $\widetilde{f}$ and $\widetilde{q}$  are real-valued functions and $\widetilde{g}^*=\widetilde{h}$. From \eqref{symmetry2}, we find that
\begin{equation*}
  \tau_{-1,0} = \widetilde{g}^*, \quad \tau_{0,-1} = \widetilde{g}.
\end{equation*}
Therefore, the bilinear equations \eqref{reduced bilinear eq1}-\eqref{reduced bilinear eq4} become
\begin{equation} \label{bilinear-2c}
   \left\{\begin{array}{l}
\left(D_{x}^{2}-4 c\right) \widetilde{f} \cdot \widetilde{f}=-4 c \widetilde{g}\, \widetilde{g}^{*}, \\
\left(D_{x}^{3}-D_t+3 i \kappa D_{x}^{2}-3\left(\kappa^{2}+2 c\right) D_{x}-6 i \kappa c\right) \widetilde{g} \cdot \widetilde{f}+6 i \kappa c \widetilde{q} \, \widetilde{g}=0, \\
\left(D_{x}+2 i \kappa\right) \widetilde{g} \cdot \widetilde{g}^{*}=2 \mathrm{i} \kappa \widetilde{q} \widetilde{f}.
\end{array}\right.
\end{equation}
Let
\begin{eqnarray*}
 f(x,t) = \widetilde{f}(x-6ct,t), \quad  g(x,t) = \widetilde{g}(x-6ct,t), \quad  q(x,t) = \widetilde{q}(x-6ct,t),
\end{eqnarray*}
then the system of bilinear equations \eqref{bilinear-2c} reduces to
\eqref{bilinear form-SS}, and thus we can obtain the following solution   to the Sasa-Satsuma equation \eqref{SS equation}
\begin{equation} 
    u=\frac{g}{f} e^{\mathrm{i}\left(\kappa(x-6 c t)-\kappa^{3} t\right)},
\end{equation}
where
\begin{eqnarray*}
  f(x,t) =  \tau_{0}(x-6ct,t), \quad g(x,t) =  \tau_{1}(x-6ct,t)
\end{eqnarray*}
and
\begin{equation*}
   \tau_{k} =  \tau_{k0}, \quad k = 0,1.
\end{equation*}
This completes the proof.


\section{Proof of Theorem \ref{thm-no recursive relation}}

In Theorem \ref{thm-recursive relation}, the rogue wave solutions contain differential operators that are recursively defined. In fact, we are able to avoid these recursive relations by modifying the differential operators. To achieve this, we introduce the notations
\begin{eqnarray*}
\mathcal{M}_{ij}^{kl} &=& \left(
  \begin{array}{ll}
    \mathfrak{m}_{2i-1,2j-1}^{kl} & \mathfrak{m}_{2i-1,2j}^{kl} \\
    \mathfrak{m}_{2i,2j-1}^{kl} & \mathfrak{m}_{2i,2j}^{kl} \\
  \end{array}
\right)
= \left(
\begin{array}{cccc}
   \mathcal{A}_{i1} \mathcal{B}_{j1} m_{11}^{kl} &  \mathcal{A}_{i1}  \mathcal{B}_{j2}  m_{12}^{kl}   \\
 \mathcal{A}_{i2} \mathcal{B}_{j1}  m_{21}^{kl} &  \mathcal{A}_{i2} \mathcal{B}_{j2}  m_{22}^{kl}   \\
\end{array}
\right)
\end{eqnarray*}
and
\begin{eqnarray*}
\left(
  \begin{array}{c}
     \varphi_{2i-1}^{kl} \\  \varphi^{kl}_{2i}
  \end{array}
\right)
 &=& \left(
\begin{array}{c}
   \mathcal{A}_{i1}   \overline{\varphi}_{1}^{kl} \\  \mathcal{A}_{i2}  \overline{\varphi}_{2}^{kl}
\end{array}
\right)
,
\quad
\left(
  \begin{array}{cc}
     \psi_{2j-1}^{kl}, \psi^{kl}_{2j}
  \end{array}
\right)
 =  \left(
\begin{array}{cc}
   \mathcal{B}_{j1} \overline{\psi}_{1}^{kl}, \mathcal{B}_{j2}  \overline{\psi}_{2}^{kl}
\end{array}
\right),
\end{eqnarray*}
where $i,j = 1,2\cdots, N$, $m^{kl}_{\alpha\beta}, \overline{\varphi}_{\alpha}^{kl},  \overline{\psi}_{\beta}^{kl}$ ($\alpha, \beta = 1,2$) are given in \eqref{matrix element-without differential operator-1}-\eqref{matrix element-without differential operator-3} and the differential operators $\mathcal{A}_{i\alpha},\mathcal{B}_{j\beta} $ are defined by
$$
 \mathcal{A}_{i\alpha}=\frac{1}{i !}\left[f_{1\alpha}(p_{\alpha}) \partial_{p_{\alpha}}\right]^{i}, \quad \mathcal{B}_{j\beta}=\frac{1}{j !}\left[f_{2\beta}(q_{\beta}) \partial_{q_{\beta}}\right]^{j}
$$
with $f_{1\alpha}(p_\alpha), f_{2\beta}(q_\beta) $  being arbitrary functions.
Since the operators $\mathcal{A}_{i\alpha}$ and $\mathcal{B}_{j\beta}$ commute with the partial differentials with respect to $x,y,t,r,s$, the functions $ \mathfrak{m}_{i j}^{kl}, \varphi_{i}^{kl}, \psi_{j}^{kl}$ would satisfy the differential and difference relations
\eqref{differential and difference relations-1}-\eqref{differential and difference relations-last 1}. Hence, for any sequence of indices $\left(i_{1}, i_{2}, \ldots, i_{N} ; j_{1}, j_{2}, \ldots, j_{N}\right)$, the determinant
\begin{eqnarray} \label{initial solutions of KP-Wp treatment}
\tau_{kl} &=&   \det_{1\leq \mu, \nu \leq N} \left(\mathcal{M}_{i_{\mu}, j_{\nu}}^{kl} \right)  
\end{eqnarray}
satisfies the bilinear equations \eqref{KP-1}-\eqref{KP-11}.

Next, by imposing the parameter relations 
\begin{equation}\label{AKP to CKP-Wp treatment}
  f_{1j} \equiv f_{2j}, \quad q_{j}=p_{j},    \quad b=-a, \quad \xi_{j 0} = \eta_{j 0}, \quad j= 1,2,
\end{equation}
which yield the reduction from AKP to CKP, and using similar arguments as in the proof of  Theorem \ref{thm-recursive relation}, we obtain
\begin{eqnarray}
\tau_{-l, -k}(x,-y, t, s, r) = \tau_{k l}(x, y, t, r, s). \label{symmetry-2-Wp treatment}
\end{eqnarray}

In what follows, we will deal with a crucial step in applying the KP hierarchy reduction method, that is, dimension reduction. In this process, several conditions are needed to accomplish this step. As a result of these conditions, the indices in the determinant \eqref{initial solutions of KP-Wp treatment} will be selected. In addition, we will  determine the functions $f_{1\alpha}, f_{2\beta}$ as well as the values of $p_{\alpha}, q_\beta$ that are involved in the matrix entries of the $\tau$ function. We also remark that the method we will utilize was named `$\mathcal{W}$-$p$ treatment' which was developed by Yang and Yang in the derivations of rogue wave solutions of the Boussinesq equation \cite{Yang2020Yang} and the three-wave equations \cite{Yang2021Yang-1}.

As shown in the proof of  Theorem \ref{thm-recursive relation},  by imposing the condition
\begin{equation} \label{dimension reduction-without recursive relation}
 \widetilde{D}_{r,s,x} \, \tau_{kl} = C  \tau_{kl},
\end{equation}
where $C$ is certain constant and $\widetilde{D}_{r,s,x}$ is the differential operator defined in \eqref{differential operator-dimension reduction}, we can reduce the bilinear equations \eqref{KP-1}-\eqref{KP-11} to \eqref{reduced bilinear eq1}-\eqref{reduced bilinear eq4}. Therefore, it suffices to find conditions under which the equation \eqref{dimension reduction-without recursive relation} holds.

 It is clear that
\begin{eqnarray*}
 \widetilde{D}_{r,s,x} \mathcal{M}_{ij}^{kl} 
 &=& \left(
\begin{array}{cccc}
   \mathcal{A}_{i1} \mathcal{B}_{j1} [F(p_{1}) + F(q_{1})] m_{11}^{kl} &  \mathcal{A}_{i1}  \mathcal{B}_{j2} [F(p_{1}) + F(q_{2})]  m_{12}^{kl}   \\
 \mathcal{A}_{i2} \mathcal{B}_{j1} [F(p_{2}) + F(q_{1})]  m_{21}^{kl} &  \mathcal{A}_{i2} \mathcal{B}_{j2} [F(p_{2}) + F(q_{2})]  m_{22}^{kl}   \\
\end{array}
\right)
\end{eqnarray*}
where
\begin{equation*}
  F(p) = \dfrac{1}{p-a} + \dfrac{1}{p+a} - \dfrac{p}{c}.
\end{equation*}
Inspired by the work of Yang and Yang \cite{Yang2020Yang,Yang2021Yang-1},
we rewrite the functions $f_{1\alpha}(p_{\alpha})$ and $f_{2\beta}(q_{\beta}) $ in the form
\begin{equation} \label{relation between f and W}
  f_1(p_\alpha) =   \dfrac{\mathcal{U}_{\alpha}(p_\alpha)}{ \mathcal{U}'_{\alpha}(p_\alpha)},  \quad   f_2(q_\beta) =    \dfrac{\mathcal{V}_{\beta}(q_\beta)  }{ \mathcal{V}'_{\beta}(q_\beta) },
\end{equation}
where the new variables $\mathcal{U}_{\alpha}$  and $\mathcal{V}_{\beta} $ are introduced such that
\begin{equation*}
 f_{\alpha} \partial_{p_\alpha} = \partial_{\ln } \mathcal{U}_{\alpha}, \quad  f_{\beta} \partial_{q_\beta} = \partial_{\ln } \mathcal{V}_{\beta}.
\end{equation*}
The reason for this representation of $f_{1\alpha}(p_{\alpha})$ and $f_{2\beta}(q_{\beta}) $ is as follows. With the above relations, we may get (see \cite{Yang2021Yang-1})
\begin{equation}
  \mathcal{A}_{i\alpha} F(p_\alpha)=\sum_{k=0}^{i} \frac{1}{k !}\left[\left(f_{1\alpha} \partial_{p_\alpha}\right)^{k} F(p_\alpha)\right] \mathcal{A}_{i-k}, \quad \alpha = 1,2,
\end{equation}
and
\begin{equation}
  \mathcal{B}_{j\beta} F(q_\beta)=\sum_{l=0}^{i} \frac{1}{l !}\left[\left(f_{1\beta} \partial_{q_\beta}\right)^{l} F(q_\beta)\right] \mathcal{B}_{j-l}, \quad \beta = 1,2.
\end{equation}
Consequently, it follows that
\begin{eqnarray*}
 \widetilde{D}_{r,s,x} \mathcal{M}_{ij}^{kl}
  &=& \left(
  \begin{array}{ll}
    \mathfrak{a}_{2i-1,2j-1}^{kl} & \mathfrak{a}_{2i-1,2j}^{kl} \\
    \mathfrak{a}_{2i,2j-1}^{kl} & \mathfrak{a}_{2i,2j}^{kl} \\
  \end{array}
\right)
\end{eqnarray*}
where
\begin{eqnarray}
&& \mathfrak{a}_{2i-1,2j-1}^{kl} = \displaystyle{  \sum_{m=0}^{i}} \frac{1}{m !}\left[\left(f_{11} \partial_{p_1}\right)^{m} F(p_1)\right] \mathfrak{m}_{2(i-m)-1 , 2j-1}^{kl} + \displaystyle{  \sum_{n=0}^{j}} \frac{1}{n !}\left[\left(f_{21} \partial_{q_1}\right)^{n} F(q_1)\right] \mathfrak{m}_{2i-1 , 2(j-n)-1}^{kl} \qquad \label{entry-dimension reduction-without recursive relation-1}
 \\
&&   \mathfrak{a}_{2i-1,2j}^{kl}  = \displaystyle{  \sum_{m=0}^{i}} \frac{1}{m !}\left[\left(f_{11} \partial_{p_1}\right)^{m} F(p_1)\right] \mathfrak{m}_{2(i-m)-1 , 2j}^{kl} + \displaystyle{  \sum_{n=0}^{j}} \frac{1}{n !}\left[\left(f_{22} \partial_{q_2}\right)^{n} F(q_2)\right] \mathfrak{m}_{2i-1 , 2(j-n)}^{kl}
 \\
&&     \mathfrak{a}_{2i,2j-1}^{kl}  = \displaystyle{  \sum_{m=0}^{i}} \frac{1}{m !}\left[\left(f_{12} \partial_{p_2}\right)^{m} F(p_2)\right] \mathfrak{m}_{2(i-m) , 2j-1}^{kl} + \displaystyle{  \sum_{n=0}^{j}} \frac{1}{n !}\left[\left(f_{21} \partial_{q_1}\right)^{n} F(q_1)\right] \mathfrak{m}_{2i , 2(j-n)-1}^{kl}
 \\
 &&  \mathfrak{a}_{2i,2j}^{kl}=  \displaystyle{  \sum_{m=0}^{i}} \frac{1}{m !}\left[\left(f_{12} \partial_{p_2}\right)^{m} F(p_2)\right] \mathfrak{m}_{2(i-m) , 2j}^{kl} + \displaystyle{  \sum_{n=0}^{j}} \frac{1}{n !}\left[\left(f_{22} \partial_{q_2}\right)^{n} F(q_2)\right] \mathfrak{m}_{2i , 2(j-n)}^{kl}. \label{entry-dimension reduction-without recursive relation-4}
\end{eqnarray}

According to the work of Ohta and Yang \cite{Ohta2012Yang}, in order to realize the dimensional reduction condition \eqref{dimension reduction-without recursive relation}, we can choose proper functions $f_{1\alpha}, f_{2\beta}$ and the values of $p_{\alpha}, q_\beta$ such that the terms with odd $m$ or $n$ on the right-hand side of equations \eqref{entry-dimension reduction-without recursive relation-1}-\eqref{entry-dimension reduction-without recursive relation-4} vanish. It turns out that this can be guaranteed if we choose $\xi,\widetilde{\xi},\eta$ and $\widetilde{\eta}$ such that
\begin{equation}\label{parameter values-without recursive relation}
  F'(\xi) = 0, \quad F'(\widetilde{\xi}) = 0,  \quad F'(\eta) = 0, \quad F'(\widetilde{\eta}) = 0
\end{equation}
and impose the conditions
\begin{equation}\label{function selection-without recursive relation}
 \left(f_{1\alpha} \partial_{p_\alpha}\right)^{2} F(p_\alpha) = F(p_\alpha), \quad \left(f_{2\beta} \partial_{q_\beta}\right)^{2} F(q_\beta) = F(q_\beta), \quad \alpha,\beta=1,2.
\end{equation}
With these conditions, we find that
\begin{eqnarray}
&& \mathfrak{a}_{2i-1,2j-1}^{kl}\left|_{\substack{p_1= \xi, q_1=\eta  \\ p_2=\widetilde{\xi}, q_2= \widetilde{\eta}}} \right. \nonumber
\\
&=& F(p_1) \displaystyle{  \sum_{m=0 \atop m: \text { even }}^{i}} \frac{1}{m !} \mathfrak{m}_{2(i-m)-1 , 2j-1}^{kl}\left|_{\substack{p_1= \xi, q_1=\eta  \\ p_2=\widetilde{\xi}, q_2= \widetilde{\eta}}} \right. + F(q_1) \displaystyle{  \sum_{n=0 \atop n: \text { even }}^{j}} \frac{1}{n !} \mathfrak{m}_{2i-1 , 2(j-n)-1}^{kl}\left|_{\substack{p_1= \xi, q_1=\eta  \\ p_2=\widetilde{\xi}, q_2= \widetilde{\eta}}} \right. \qquad \label{entry-dimension reduction-without recursive relation-evaluation-1}
 \\
&& \mathfrak{a}_{2i-1,2j}^{kl}\left|_{\substack{p_1= \xi, q_1=\eta  \\ p_2=\widetilde{\xi}, q_2= \widetilde{\eta}}} \right. \nonumber
\\
&=& F(p_1) \displaystyle{  \sum_{m=0 \atop m: \text { even }}^{i}} \frac{1}{m !} \mathfrak{m}_{2(i-m)-1 , 2j}^{kl}\left|_{\substack{p_1= \xi, q_1=\eta  \\ p_2=\widetilde{\xi}, q_2= \widetilde{\eta}}} \right. + F(q_2) \displaystyle{  \sum_{n=0 \atop n: \text { even }}^{j}} \frac{1}{n !} \mathfrak{m}_{2i-1 , 2(j-n)}^{kl}\left|_{\substack{p_1= \xi, q_1=\eta  \\ p_2=\widetilde{\xi}, q_2= \widetilde{\eta}}} \right. \qquad
 \\
&& \mathfrak{a}_{2i,2j-1}^{kl}\left|_{\substack{p_1= \xi, q_1=\eta  \\ p_2=\widetilde{\xi}, q_2= \widetilde{\eta}}} \right. \nonumber
\\
&=& F(p_2) \displaystyle{  \sum_{m=0 \atop m: \text { even }}^{i}} \frac{1}{m !} \mathfrak{m}_{2(i-m) , 2j-1}^{kl}\left|_{\substack{p_1= \xi, q_1=\eta  \\ p_2=\widetilde{\xi}, q_2= \widetilde{\eta}}} \right. + F(q_1) \displaystyle{  \sum_{n=0 \atop n: \text { even }}^{j}} \frac{1}{n !} \mathfrak{m}_{2i , 2(j-n)-1}^{kl}\left|_{\substack{p_1= \xi, q_1=\eta  \\ p_2=\widetilde{\xi}, q_2= \widetilde{\eta}}} \right. \qquad
 \\
&& \mathfrak{a}_{2i,2j}^{kl}\left|_{\substack{p_1= \xi, q_1=\eta  \\ p_2=\widetilde{\xi}, q_2= \widetilde{\eta}}} \right. \nonumber
\\
&=& F(p_2) \displaystyle{  \sum_{m=0 \atop m: \text { even }}^{i}} \frac{1}{m !} \mathfrak{m}_{2(i-m), 2j}^{kl}\left|_{\substack{p_1= \xi, q_1=\eta  \\ p_2=\widetilde{\xi}, q_2= \widetilde{\eta}}} \right. + F(q_2) \displaystyle{  \sum_{n=0 \atop n: \text { even }}^{j}} \frac{1}{n !} \mathfrak{m}_{2i , 2(j-n)}^{kl}\left|_{\substack{p_1= \xi, q_1=\eta  \\ p_2=\widetilde{\xi}, q_2= \widetilde{\eta}}} \right. .\qquad \label{entry-dimension reduction-without recursive relation-evaluation-4}
\end{eqnarray}
Then, by restricting the indices of the general determinant \eqref{initial solutions of KP-Wp treatment} to
the determinant
\begin{eqnarray} \label{final tau function-without recursive relation}
\tau_{kl} &=&   \det_{1\leq i,j\leq N} \left(\mathcal{M}_{2i-1, 2j-1}^{kl}\left|_{\substack{p_1= \xi, q_1=\eta  \\ p_2=\widetilde{\xi}, q_2= \widetilde{\eta}}} \right.  \right)  
\end{eqnarray}
and using the relations \eqref{entry-dimension reduction-without recursive relation-evaluation-1}-\eqref{entry-dimension reduction-without recursive relation-evaluation-4}, a similar argument as in the proof of Theorem \ref{thm-recursive relation} yields
\begin{equation}
  \widetilde{D}_{r,s,x} \tau_{kl} = \left[F(\xi) + F(\eta)+F(\widetilde{\xi}) + F(\widetilde{\eta})\right]N\tau_{kl}.
\end{equation}
Therefore, the function $\tau_{kl}$ given in \eqref{final tau function-without recursive relation} satisfies  the dimensional reduction condition \eqref{dimension reduction-without recursive relation} as long as \eqref{parameter values-without recursive relation} and \eqref{function selection-without recursive relation} hold.

Now we move to determine the values of $\xi,\widetilde{\xi},\eta, \widetilde{\eta}$ and the expressions of $f_{1\alpha}, f_{2\beta}$ such that they satisfy \eqref{parameter values-without recursive relation} and \eqref{function selection-without recursive relation}. From \eqref{AKP to CKP-Wp treatment}, we can get $\xi = \eta$ and $\widetilde{\xi} = \widetilde{\eta}$. Similar to the proof of Theorem \ref{thm-recursive relation}, we should require $\widetilde{\xi}  = \xi^*$ and $\widetilde{\eta}  = \eta^*$ to make sure that the complex conjugate condition is fulfilled. On the other hand, the terms $1/(p_\alpha+q_\beta)$ in \eqref{matrix element-without differential operator-1} indicate that $\xi + \widetilde{\eta} \not = 0 $ and $  \widetilde{\xi} + \eta \not = 0 $ while the expressions of $\tau_{kl}$ in \eqref{final tau function-without recursive relation} show that $\xi \not = \widetilde{\xi}$. Therefore $\xi^2$ cannot be real. Since the equations  \eqref{parameter p} and \eqref{parameter values-without recursive relation} are equivalent, from Remark \ref{root sturcture}, we find that this occurs if and only if
 $$c < 0, \quad c + 4 \kappa ^2>0.$$
In this case, the roots of the equation \eqref{parameter p} demonstrate a symmetric structure (see Remark \ref{root sturcture}). Hence, no matter which root we choose, we get the same solution. Apart from this, the symmetric structure indicates that all the roots of $F'(p) = 0$ are simple.

In order to find $f_{11}$, it suffices to solve the differential equation
\begin{equation}
 \left(f_{11} \partial_{p_1}\right)^{2} F(p_1) = F(p_1).
\end{equation}
With \eqref{relation between f and W}, it becomes
\begin{equation} 
\partial_{\ln\mathcal{U}_{1} }^{2} F(p_1)=F(p_1). 
\end{equation}
Normalize $\mathcal{U}_{1}(\xi) =   1, $ then the unique solution to this second-order differential equation with the initial condition $F(\xi) = F(\xi), F'(\xi) = 0$ is
\begin{equation} \label{relation between p and W}
  F(p_1)=\frac{1}{2} F(\xi)\left(\mathcal{U}_{1}(p_1)+\frac{1}{\mathcal{U}_{1}(p_1)}\right).
\end{equation}
Hence, with  \eqref{relation between f and W}, it follows that
\begin{equation} \label{U1}
 \mathcal{U}_{1}(p_1)=\frac{F(p_1) \pm \sqrt{F^{2}(p_1)-F^{2}\left(\xi\right)}}{F\left(\xi\right)}
\end{equation}
which gives
\begin{equation} \label{solution of f11}
  f_{11}(p_1)=\pm \frac{\sqrt{F^{2}(p_1)-F^{2}\left(\xi\right)}}{F^{\prime}(p_1)}.
\end{equation}
By changing $(f_{11}, p_1, \xi)$ to $(f_{12}, p_2, \xi^*)$, $(f_{21}, q_1, \xi)$ or $(f_{22}, q_2, \xi^*)$, we may solve for  $f_{12}$, $f_{21}$ and $f_{22}$ as well, which share the same form as \eqref{solution of f11}. This completes the dimension reduction procedure. As a consequence, the bilinear equations \eqref{KP-1}-\eqref{KP-11} are reduced to \eqref{reduced bilinear eq1}-\eqref{reduced bilinear eq4} and hence, we can take $y=r=s=0$ because the bilinear equations \eqref{reduced bilinear eq1}-\eqref{reduced bilinear eq4} do not involve derivatives with respect to $y,r$ and $s$.
Then according to \eqref{symmetry-2-Wp treatment}, we have
\begin{equation} \label{symmetry2-Wp treatment}
\tau_{k l}(x, t)=\tau_{-l, -k}(x, t).
\end{equation}

Now we consider the complex conjugate reduction. To this end, we set $a$ to be purely imaginary and impose the parameter constraint
\begin{equation}
 \xi_{1 0} = \xi_{2 0}^*.
\end{equation}
In view of $[F(p)]^* \equiv F(p^*)$, we find $[F^{(n)}(p)]^* \equiv F^{(n)}(p^*)$ holds for any positive integer $n$. This implies that
\begin{equation} 
  [f_{11}(p)]^* \equiv f_{12}(p^*), \quad [f_{21}(p)]^* \equiv f_{22}(p^*), \quad
  [f_{11}^{(n)}(p)]^* \equiv f_{12}^{(n)}(p^*), \quad [f_{21}^{(n)}(p)]^* \equiv f_{22}^{(n)}(p^*),
\end{equation}
where $n$ is a positive integer. It then follows that
\begin{eqnarray*}
\left(\mathcal{M}_{ij}^{kl}\right)^*\left|_{\substack{p_1 = q_1 = \xi  \\ p_2 = q_2 =\xi^*}} \right. &=& \left(
  \begin{array}{cc}
  \left(  \mathfrak{m}_{2i-1,2j-1}^{kl}\right)^* &  \left(\mathfrak{m}_{2i-1,2j}^{kl}\right)^* \\
     \left(\mathfrak{m}_{2i,2j-1}^{kl}\right)^* &  \left(\mathfrak{m}_{2i,2j}^{kl}\right)^* \\
  \end{array}
\right) \left|_{\substack{p_1 = q_1 = \xi  \\ p_2 = q_2 =\xi^*}} \right.
\\
&=&  \left(
  \begin{array}{ll}
    \mathfrak{m}_{2i,2j}^{lk} & \mathfrak{m}_{2i,2j-1}^{lk} \\
    \mathfrak{m}_{2i-1,2j}^{lk} & \mathfrak{m}_{2i-1,2j-1}^{lk} \\
  \end{array}
\right) \left|_{\substack{p_1 = q_1 = \xi  \\ p_2 = q_2 =\xi^*}} \right. .
\end{eqnarray*}
A similar argument as in the proof of Theorem \ref{thm-recursive relation} gives
\begin{equation}
\tau^*_{ k l } =  \tau_{l k}.
\end{equation}
Define
$$
\widetilde{f}=\tau_{00}, \quad \widetilde{g}=\tau_{10}, \quad \widetilde{h}=\tau_{01}, \quad \widetilde{q}=\tau_{11},
$$
then we find that   $\widetilde{f}$ and $\widetilde{q}$  are real-valued functions and $\widetilde{g}^*=\widetilde{h}$. With \eqref{symmetry2-Wp treatment}, we get
\begin{equation*}
  \tau_{-1,0} = \widetilde{g}^*, \quad \tau_{0,-1} = \widetilde{g}.
\end{equation*}
Therefore, the bilinear equations \eqref{reduced bilinear eq1}-\eqref{reduced bilinear eq4} are reduced to \eqref{bilinear-2c}.
Let
\begin{eqnarray*}
 f(x,t) = \widetilde{f}(x-6ct,t), \quad  g(x,t) = \widetilde{g}(x-6ct,t), \quad  q(x,t) = \widetilde{q}(x-6ct,t),
\end{eqnarray*}
then $f,g$ and $g$ satisfy the bilinear equations \eqref{bilinear form-SS}. Hence, with $   \tau_{k} =  \tau_{k0}, k = 0,1$,  rational solutions to the Sasa-Satsuma equation \eqref{SS equation} are obtained via the transformation \eqref{transformation1}.

Finally, we apply the method proposed in \cite{Yang2020ChenYang} to introduce free parameters into these solutions. This is accomplished by choosing $\xi_{10}$ as
\begin{equation}
  \xi_{10}=\sum_{r=1}^{\infty}  a_{n}  \ln ^{n} \mathcal{U}_{1}(p_1),
\end{equation}
where $\mathcal{U}_{1}(p_1)$ is defined in \eqref{U1} and $\hat{a}_{n}$ are free complex constants. Thus the proof of Theorem \ref{thm-no recursive relation} is completed.


\section{Proof of Theorem \ref{thm-Schur polynomials}}\label{Proof of Theorem-Schur polynomials}

In order to represent the rational solutions of the Sasa-Satsuma equation derived in Theorem \ref{thm-no recursive relation} in terms of Schur polynomials, we first consider the generator of the differential operators $\left[f_{1\alpha}(p_{\alpha}) \partial_{p_{\alpha}}\right]^{i} \left[f_{2\beta}(q_{\beta}) \partial_{q_{\beta}}\right]^{j}$ given by
\begin{equation}
\mathcal{G}=\sum_{i=0}^{\infty}\sum_{j=0}^{\infty}\frac{\kappa^i}{i!}\frac{\lambda^j}{j!}(f_{1\alpha}\partial_{p_{\alpha}})^i(f_{2\beta}\partial_{q_{\beta}})^j. 
\end{equation}
Using the relations \eqref{relation between f and W}, the above generator can be rewritten as
\begin{equation*}
  \mathcal{G}=\sum_{i=0}^{\infty} \sum_{j=0}^{\infty} \frac{\kappa^{i}}{i !} \frac{\lambda^{j}}{j !}\left(\partial_{\ln  \mathcal{U}_\alpha}\right)^{i}\left(\partial_{\ln \mathcal{V}_\beta} \right)^{j}=\exp \left(\kappa \partial_{\ln \mathcal{U}_\alpha}+\lambda \partial_{\ln \mathcal{V}_\beta}\right).
\end{equation*}
This implies that, for any function $H(X,Y)$, we have
\begin{equation}
\mathcal{G}H(X,Y)=H(e^\kappa X,e^\lambda Y).
\end{equation}
From \eqref{relation between p and W}, we see that $p_{\alpha}$ and $q_{\beta}$ are related to $\mathcal{U}_\alpha$ and $\mathcal{V}_\beta$ respectively and hence
\begin{equation*}
 p_{\alpha} = p_{\alpha}(\mathcal{U}_\alpha), \quad q_{\beta}= q_{\beta} (\mathcal{V}_\beta).
\end{equation*}
In addition, from the proof of Theorem \ref{thm-no recursive relation}, we may deduce that
\begin{equation}
  \mathcal{U}_1 (\xi) = \mathcal{U}_2 ( \xi^* ) = \mathcal{V}_1 (\xi) = \mathcal{V}_2 (\xi^*)  = 1.
\end{equation}
Therefore, by denoting
\begin{equation} 
  \mathfrak{p_\alpha} (\kappa) = p_\alpha(e^{\kappa}), \quad \mathfrak{q}_\beta (\lambda) = q_\beta(e^{\lambda})
\end{equation}
and acting $\mathcal{G}$ on $ m_{\alpha,\beta}^{k}$ defined in Theorem \ref{thm-no recursive relation}, we can get
\begin{eqnarray}
\mathcal{G} m_{\alpha,\beta}^{k}\left|_{\substack{p_1 = q_1 = \xi  \\ p_2 = q_2 =\xi^*}} \right. &=&\frac{1}{\mathfrak{p_\alpha} (\kappa)+\mathfrak{q}_\beta (\lambda)}\left(-\dfrac{\mathfrak{p_\alpha} (\kappa)-a}{\mathfrak{q}_\beta (\lambda)+a}\right)^{k} \exp \left[\sum_{n=1}^{\infty}\left(\hat{a}_{n\alpha} \kappa^{n}+\hat{a}_{n\beta} \lambda^{n}\right)\right] \nonumber
 \\
&& \exp\left[\left(\mathfrak{p_\alpha} (\kappa)+\mathfrak{q}_\beta (\lambda)\right) x + \left(\mathfrak{p^3_\alpha} (\kappa)+\mathfrak{q}^3_\beta (\lambda)\right)t \right],  \label{differential operator-Schur polynomial}
\end{eqnarray}
where
\begin{equation*}
   \hat{a}_{n1} =  a_{n}  , \quad  \hat{a}_{n2} =  a_{n}^*.
\end{equation*}
Denote by $\mathfrak{p} (\kappa)= \mathfrak{p}_1 (\kappa),\mathfrak{q} (\lambda)= \mathfrak{q}_1 (\lambda)$, then it follows from \eqref{parameter relation-recursive relation} that
\begin{equation*}
\mathfrak{p}_2 (\kappa) = \mathfrak{p}^* (\kappa), \quad \mathfrak{q} (\lambda)  = \mathfrak{p} (\lambda), \quad \mathfrak{q}_2 (\lambda)  = \mathfrak{p}^* (\lambda),
\end{equation*}
where $\kappa,\lambda$ are real variables.
With \eqref{differential operator-Schur polynomial}, we have
\begin{eqnarray}
&&\frac{1}{m_{\alpha,\beta}^{k}}\mathcal{G} m_{\alpha,\beta}^{k}\left|_{\substack{p_1 = q_1 = \xi  \\ p_2 = q_2 =\xi^*}} \right.  \nonumber
\\
&=&\frac{p_{\alpha}+q_{\beta}}{\mathfrak{p_\alpha} (\kappa)+\mathfrak{q}_\beta (\lambda)}\left(\dfrac{\mathfrak{p_\alpha} (\kappa)-a}{p_{\alpha}-a}\right)^{k} \left(\dfrac{\mathfrak{q}_\beta (\lambda)+a}{q_{\beta}+a}\right)^{-k} \exp \left[\sum_{n=1}^{\infty}\left(\hat{a}_{n\alpha} \kappa^{n}+\hat{a}_{n\beta} \lambda^{n}\right)\right] \nonumber
 \\
 &&\exp\left[\left(\mathfrak{p_\alpha} (\kappa)- p_\alpha+\mathfrak{q}_\beta (\lambda) -q_\beta\right) x + \left(\mathfrak{p^3_\alpha} (\kappa)- p_\alpha^3+\mathfrak{q}^3_\beta(\lambda)-q_\beta^3\right)  t  \right] \left|_{\substack{p_1 = q_1 = \xi  \\ p_2 = q_2 =\xi^*}} \right. . \label{general differential operator-Schur polynomial}
\end{eqnarray}
In order to connect the right hand side of the above equation with Schur polynomials, we have to expand it as a series in $\kappa$ and $\lambda .$



  Without loss of generality, we may assume $\alpha=\beta=1$ and denote by
\begin{eqnarray}
   \hat{p}_1 = \mathfrak{p}'(0), \quad    \hat{q}_1 = \mathfrak{q}'(0). 
\end{eqnarray}
Note that when $(\kappa, \lambda)$ is around $(0,0)$, we have \cite{Yang2020Yang}
\begin{eqnarray}
 \frac{1}{g(\kappa)+h(\lambda)}&=&\frac{g(0)+h(0)}{[g(\kappa)+h(0)][h(\lambda)+g(0)]} \frac{1}{1-\dfrac{g(\kappa)-g(0)}{g(\kappa)+h(0)} \dfrac{h(\lambda)-h(0)}{h(\lambda)+g(0)}} \\
 &=&\frac{g(0)+h(0)}{[g(\kappa)+h(0)][h(\lambda)+g(0)]}  \sum_{\nu=0}^{\infty}\left[\frac{g(\kappa)-g(0)}{g(\kappa)+h(0)} \frac{h(\lambda)-h(0)}{h(\lambda)+g(0)}\right]^{\nu}.
\end{eqnarray}
  Using the above identity, the first term on the right hand side of \eqref{general differential operator-Schur polynomial} can be rewritten as
  \begin{eqnarray}
\frac{2\xi}{\mathfrak{p}+\mathfrak{q} } 
&=&
\frac{4\xi^{2}}{\left(\mathfrak{p}+\xi\right)\left(\mathfrak{q} +\xi\right)} \sum_{\nu=0}^{\infty}\left[\frac{\left(\mathfrak{p}-\xi\right)\left(\mathfrak{q} -\xi\right)}{\left(\mathfrak{p}+\xi\right)\left(\mathfrak{q} +\xi\right)}\right]^{\nu} \nonumber
\\
&=&\frac{4\xi^{2}}{\left(\mathfrak{p}+\xi\right)\left(\mathfrak{q} +\xi\right)} \sum_{\nu=0}^{\infty}\left(\frac{\hat{p}_{1} \hat{q}_{1}}{4\xi^{2}} \kappa \lambda\right)^{\nu}\left(\frac{2 \xi}{\hat{p}_{1} \kappa} \frac{\mathfrak{p}-\xi}{\mathfrak{p}+\xi}\right)^{\nu}\left(\frac{2\xi}{\hat{q}_{1} \lambda}  \nonumber \frac{\mathfrak{q}-\xi}{\mathfrak{q}+\xi}\right)^{\nu}
\\
&=&\sum_{\nu=0}^{\infty}\left(\frac{\hat{p}_{1} \hat{q}_{1}}{4\xi^{2}} \kappa \lambda\right)^{\nu} \exp \left(\sum_{n=1}^{\infty}\left(\nu s_{n}-b_{n}\right) \kappa^{n}+\left(\nu s_{n}-b_{n}\right) \lambda^{n}\right) \label{cofactor}
  \end{eqnarray}
  while the remaining terms on the right hand side of \eqref{general differential operator-Schur polynomial} can be rewritten as
\begin{eqnarray} \label{exponential term}
\exp\left\{\sum_{n=1}^{\infty} \kappa^n (c_n x + d_n t + k h^+_n) + \sum_{n=1}^{\infty} \lambda^n (c_n x + d_n t - k h^-_n) + \sum_{n=1}^{\infty}\left( a_{n} \kappa^{n}+ a_{n}  \lambda^{n}\right)  \right\}. 
\end{eqnarray}
Then substituting \eqref{cofactor} and \eqref{exponential term} into \eqref{general differential operator-Schur polynomial} yields
\begin{eqnarray}
\frac{1}{m_{11}^{k}}\mathcal{G} m_{11}^{k}\left|_{\substack{p_1 = q_1 = \xi  \\ p_2 = q_2 =\xi^*}} \right. &=& \sum_{\nu=0}^{\infty}\left(\frac{\hat{p}_{1} \hat{q}_{1}}{4\xi^{2}} \kappa \lambda\right)^{\nu}  \nonumber
\\
&&
 \exp \left(\sum_{n=1}^{\infty}\left(x_{n} + \nu s_{n} - b_n + k h_n^+ \right) \kappa^{n}+\left(x_{n} + \nu s_{n} - b_n - k h_n^- \right) \lambda^{n}\right), \nonumber
\end{eqnarray}
Comparing  the coefficients of $\kappa^i\lambda^j$ on both sides of the above equation, it follows that

 \begin{equation*}
   \frac{ \mathcal{A}_{i1} \mathcal{B}_{j1}m_{11}^{(k, l)}}{(p_1+q_1)m_{11}^{(k, l)}} \left|_{\substack{p_1 = q_1 = \xi  \\ p_2 = q_2 =\xi^*}} \right. =\sum_{\nu=0}^{\min (i, j)} \frac{1}{2\xi} \left(\frac{\hat{p}_{1}^2}{4\xi^{2}}\right)^{\nu} S_{i-\nu}\left(\mathbf{x}+\nu \mathbf{s} - \mathbf{b} + k\mathbf{h}^+\right) S_{j-\nu}\left(\mathbf{x}+\nu \mathbf{s} - \mathbf{b} - k\mathbf{h}^-\right),
 \end{equation*}
 where $\mathcal{A}_{i1},\mathcal{B}_{j1} $ are defined in \eqref{differential operator-without recursive relation}. Using similar arguments, it can be shown that
 \begin{eqnarray*}
       \frac{ \mathcal{A}_{i1} \mathcal{B}_{j2}m_{12}^{(k, l)}}{(p_1+q_2)m_{12}^{(k, l)}} \left|_{\substack{p_1 = q_1 = \xi  \\ p_2 = q_2 =\xi^*}} \right. &=& \sum_{\nu=0}^{\min (i, j)} \frac{1}{\xi+\xi^*} \left(\frac{|\hat{p}_{1}|^2}{(\xi+\xi^*)^{2}}\right)^{\nu}
       \\
       && S_{i-\nu}\left(\mathbf{x}+\nu \mathbf{\hat{s}}+k \mathbf{h}^{+}- \mathbf{\hat{b}}\right) S_{j-\nu}\left(\mathbf{x}^{*}+\nu \mathbf{\hat{s}}^*-k(\mathbf{h}^+)^* - \mathbf{\hat{b}}^*\right) \nonumber
       \\
      \frac{ \mathcal{A}_{i2} \mathcal{B}_{j1}m_{21}^{(k, l)}}{(p_2+q_1)m_{12}^{(k, l)}} \left|_{\substack{p_1 = q_1 = \xi  \\ p_2 = q_2 =\xi^*}} \right. &=& \sum_{\nu=0}^{\min (i, j)} \frac{1}{\xi^*+\xi} \left(\frac{|\hat{p}_{1}|^2}{(\xi^*+\xi)^{2}}\right)^{\nu}
       \\
       && S_{i-\nu}\left(\mathbf{x}^*+\nu \mathbf{\hat{s}}^*+k (\mathbf{h}^{-})^* - \mathbf{\hat{b}}^*\right) S_{j-\nu}\left(\mathbf{x}+\nu \mathbf{\hat{s}}-k\mathbf{h}^- - \mathbf{\hat{b}}\right) \nonumber
       \\
       \frac{ \mathcal{A}_{i2} \mathcal{B}_{j2}m_{22}^{(k, l)}}{(p_2+q_2)m_{22}^{(k, l)}} \left|_{\substack{p_1 = q_1 = \xi  \\ p_2 = q_2 =\xi^*}} \right. &=& \sum_{\nu=0}^{\min (i, j)} \frac{1}{2 \xi^*} \left(\frac{\left(\hat{p}_{1}^*\right)^2}{4 (\xi^*)^{2}}\right)^{\nu}
       \\
       && S_{i-\nu}\left(\mathbf{x}^*+\nu \mathbf{s}^*+k (\mathbf{h^-})^*- \mathbf{b}^*\right) S_{j-\nu}\left(\mathbf{x}^*+\nu \mathbf{s}^* -k(\mathbf{h^+})^* - \mathbf{b}^* \right) \nonumber
 \end{eqnarray*}
  Let
 \begin{equation}
  \sigma_{k}=\frac{\tau_{k}}{\left( \displaystyle{\prod_{\alpha,\beta=1}^2} (p_\alpha+q_\beta) m_{\alpha,\beta}^{(k, l)}\left|_{\substack{p_1 = q_1 = \xi  \\ p_2 = q_2 =\xi^*}} \right.\right)^{N}}.
 \end{equation}
Considering the fact that the solution set of \eqref{SS equation} is invariant under multiplication of $e^{i\theta}$, where $\theta$ is real, the function \eqref{rogue wave solution-Schur polynomial}
still satisfies the Sasa-Satsuma equation \eqref{SS equation}. In other words, the rational solutions presented in Theorem \ref{thm-no recursive relation} can be expressed in terms of Schur polynomials. Thus the proof of Theorem \ref{thm-Schur polynomials} is completed.

\section{Conclusions} \label{Conclusions}

In summary, we have derived general rogue wave solutions to the Sasa-Satsuma equation by applying the KP hierarchy reduction method. These solutions are expressed in terms of Gram-type determinants in three different forms.  The first two forms are expressed in terms of differential operators. The difference between them is that the differential operators are recursively defined for one form whereas they are no longer recursively defined for the other form.
  Instead of using differential operators, the third form is represented by Schur polynomials based on the solutions involving differential operators without recursive relations. Owing to the complexity of the Sasa-Satsuma equation, it is shown that its solution structure is more complicated compared with many other integrable equations including the NLS equation which can be reduced from the Sasa-Satsuma equation. As such, we propose a kernel of $2\times 2$ matrix in the Gram-type determinant solutions of the bilinear equations in the KP hierarchy. To the best of knowledge, this solution form has not been reported before. In addition,  the intermediate computations are much more involved compared with most of the integrable equations that can be solved by the same method due to the multiple corresponding bilinear equations in the KP hierarchy.

\section*{Acknowledgements}

 B.F. Feng was partially supported by National Science Foundation (NSF) under Grant No. DMS-1715991 and U.S. Department of Defense (DoD), Air Force for Scientific Research (AFOSR) under grant No. W911NF2010276.
C.F. Wu was supported by the National Natural Science Foundation of China (Grant Nos. 11701382 and 11971288) and Guangdong Basic and Applied Basic Research Foundation, China (Grant No. 2021A1515010054).

\end{document}